\documentclass[12pt]{article} 

\usepackage{graphicx}
\usepackage{amsmath}
\usepackage{amssymb}
\usepackage{hyperref}
\usepackage[height=8.8in,width=6.45in]{geometry}
\usepackage{tikz}
\usepackage{cite}
\usepackage{amscd}
\usepackage{setspace}
\usepackage{bbm}
\usepackage{dsfont}
\usepackage{cancel}
\usepackage{relsize}
\usepackage[utf8]{inputenc}
\usepackage{xcolor}
\usepackage{tabularx}

\numberwithin{equation}{section}

\begin{document}
\begin{titlepage}

\begin{flushright}

\end{flushright}

\vskip 3cm

\begin{center}
{\Large \bf
Heisenberg Spin Chain And Supersymmetric Gauge Theory}

\vskip 2.0cm

\ Wei Gu \\

\bigskip

\begin{tabular}{cc}
 \textit{Max-Planck-Institute f\"{u}r Mathematik, Vivatsgasse 7, D-53111 Bonn, Germany}\\
 \textrm{and}\\
\textit{Department of Physics, Virginia Tech, 850 West Campus Dr.}\\
 \textit{ Blacksburg, VA 24061, USA}\\

\end{tabular}

\vskip 1cm

 {\tt guwei@mpim-bonn.mpg.de }

\vskip 1cm

\textbf{Abstract}

\end{center}

\medskip
\noindent

We show how a Heisenberg spin chain emerges from the two-dimensional ${\cal N}$=(2,2) gauge theory at an intermediate scale, which relies on the renormalization group flow guided by the global symmetries and the dynamics of domain walls. The discussion of higher-dimensional gauge theories with four supercharges is similar by compactifying them into two dimensions. Instead of utilizing the bosonic fields solely in the literature, adopting the fermionic degrees of freedom of gauge theory is a crucial step in our construction. From the perspective of the spin chain, we observe that the Seiberg(-like) duality between two gauge theories is manifest since it is a finite symmetry in the closed spin chain. Based on this, we further conjecture gauge theories with the same global symmetries could be unified into a single system: the Heisenberg spin chain. At least, we prove this conjecture at the intermediate scale. Finally, we also comment on other formulations of an integrable system centered on the Heisenberg spin chain.

\bigskip
\vfill
\begin{center}
\textit{To brave people}
\end{center}
\end{titlepage}

\setcounter{tocdepth}{2}
\tableofcontents
\section{Introduction}
The folklore, a quantum field theory is a quantum mechanics with infinite degrees of freedom, is too rough. There are new phenomena in quantum field theory that can not descend into quantum mechanics. For example, the Higgs mechanism can only happen in quantum field theory. To see this, consider a system with multiple vacua. Besides the usual perturbative fluctuation, it also includes the non-perturbative degrees of freedom connecting two different vacua. And they are domain walls. Its amplitude can, in general, only be suppressed at a large spatial volume limit\footnote{In cases such as 1+d quantum field theory with an emergent d-form symmetry at a finite scale, the suppression of domain walls, charged under the d-form symmetry, can happen at that scale\cite{Gu:2021beo}. }, where the quantum field theory localizes at a specific vacuum which is exactly what the Higgs mechanism tells us. On the other hand, the domain wall in quantum mechanics is an instanton. So we have no way to ignore its amplitude unless in the situation where the energy barrier between two different vacua is infinite.

The profound connections between quantum field theory and quantum mechanics deserve further investigation. In this paper, we want to convince the readers that the Heisenberg spin chain as a quantum mechanical model\footnote{It is often advertised as a statistical mechanical model. In this paper, we focus on its quantum mechanical behavior.} would emerge from the supersymmetric gauge theory at an intermediate scale. Heisenberg spin chain \footnote{The initial spin chain studied by Heisenberg is the so-called $SU(2)$ XXX model. In this paper, the Heisenberg spin chain could represent other models as well.} is a 1-dimensional lattice with $N$ sites, where at each point of a lattice, a spin $s_{i}\in\{\pm\}$ represents the local degrees of freedom with spin either up or down. Traditionally, it associates with a Hamiltonian that describes the interactions that only occur between two nearest-neighbor sites. However, a larger state space and a more generic interaction could be allowed. Despite the simple setup, it is an important model used in the study of critical points and phase transitions of magnetic systems. And it is also related to the prototypical Ising model.

On the contrary, the structure of quantum field theory is usually extraordinarily complicated, although it has been shown successfully in many areas of theoretical physics and mathematics. One salient concept in quantum field theory is that physics changes by varying the
energy scale, e.g., the coupling parameters depend on the scale, especially since many things are still unknown at the strong couplings. However, in a notable case, supersymmetric quantum field theory, we have more tools to tackle the strong-coupling region. Since one can use more RG-invariant quantities, such as BPS spectra, to probe the strong-coupling behaviors of a theory. A particular one is the (anti-)domain walls related to our story. They exist in theory with multiple vacua. Now, imagine mapping ground states to spin configurations of abstract $N$ sites. Then, the (anti)-domain wall that fluctuates between two ``adjacent vacua" has a natural correspondence in the spin chain: the interactions of two adjacent sites. However, several issues need to be addressed before confirming this picture. The first issue is that there is also a perturbative spectrum, besides domain walls, around each vacuum. In a mass gap theory, one may suppress them at low energies. However, here comes the second issue. The mass of a domain wall is proportional to $L^{d-1}$ where $L$ is the size of the spatial direction, and $d$ is the spatial dimension. So the domain wall is usually more suppressed in infrared even though the tension of a domain wall could be RG-protected, which is one reason for the Higgs mechanism in quantum field theory.

One can solve these two issues in two-dimensional supersymmetric quantum field theory simultaneously. To see this, we first notice that the masses of BPS domain walls in two dimensions are finite. Meanwhile, the non-BPS perturbative spectra usually become infinitely heavy in the far infrared of a mass gap theory. In this context, it does have an intermediate scale where the only remaining \textit{fundamental} dynamic objects are domain walls. The ``LSZ"-like formula due to these degrees of freedom would be extremely interesting.  One step in the usual LSZ reduction is to map the Hilbert space defined in the free theory to the one in the interaction theory. Since the dynamical objects, domain walls, are charged by a finite symmetry $G$, the interacting Hilbert space has to be neutral under the global group $G$. This Hilbert space is exactly the state space of an emergent spin chain. Finally, we want to point out that the spin chain can also emerge from a higher dimensional quantum field theory if we put it on $\mathbb{R}^{2}\times C$, where $C$ is a compact space. In this setup, the domain walls in two dimensions also depend on the isometry of $C$. Thus, this opens largely unexplored deep connections between supersymmetric gauge theory and spin chain. From the point of view of the spin chain, it may suggest that supersymmetry should appear in nature.\\

 Based on the above vision, here comes our main claim: \textit{Heisenberg spin chain emerges from an ${\cal N }$=(2,2) gauge theory on space-time $\mathbb{R}^{2}\times C$ at the intermediate scale.}\\

Several other technical points need to be mentioned to achieve this. To study the vacuum structure of a nonabelian supersymmetric gauge theory, one can go to the so-called ``generic Coulomb branch", where the gauge group $G$ will be Higgsed to a semi-direct product of its maximal torus and Weyl group. In our situation, we usually have the adjoint representation of the gauge group, a scalar $\sigma$, to label the vacua. The generic Coulomb branch says $\rho^{a}\sigma_{a}\neq 0$ and $\alpha^{a}\sigma_{a}\neq 0$ for $a \in\{1,\cdots , {\rm rank}G\}$, where $\rho$ are weights of representations and $\alpha$ are roots of $G$. This semiclassical analysis is, in fact, also correct in the exact quantum theory \cite{Gu:2018fpm}. We observe that this dynamical constraint in nonabelian gauge theory descends to the fermionic statistic feature of a spin chain. A further detailed study of the ground state wave functions says they all have a fermion number to be $k$. By assuming $N$ possible values of each $\sigma$ field, one can observe that the vacuum structure of this gauge theory looks similar to the \textit{excited} states in a spin chain. Because those states can be represented by picking up $k$ spin-up sites out of $N$ sites in a spin chain. Moreover, descending from ${\cal N}=(2,2)$ gauge theories in a complicated route, one can define four fermionic operators $\lambda_{\pm}$ and ${\bar\lambda}_{\pm}$ on each site. Hence, one can construct each operator of a spin chain by these emergent operators. For example, the spin operators on each site are composite ones: $S_{+}={\bar \lambda}_{+}\lambda_{-}$ and $S_{-}={\bar \lambda}_{-}\lambda_{+}$. These composite operators do not necessarily have corresponding field configurations in a single gauge theory since they change the rank of a gauge group under the dictionary map. However, they can make sense for a class of gauge theories, which indicates a new connection among gauge theories that can not be easily seen in field-theoretic language. By contrast, the operator ${\cal D}_{i}=S_{+,i}S_{-,i+1}$ in a spin chain keeps the gauge group unchanged, so it can be defined in a single gauge theory and represents a domain wall field configuration. Finally, guided by symmetries, one can derive the Hamiltonians of a spin chain by requiring that the ground state wave functions of a gauge theory are their eigenstates.

Global symmetries play crucially in our discoveries since they are renormalization group flow invariant. Therefore, symmetries of two-dimensional gauge theory can be described by those fermions $\lambda_{\pm,i}$ and ${\bar\lambda}_{\pm,i}$ in the low energies effective theory as well. For example, the generator of the axial R-symmetry is $F_{A}=\sum_{i}\left({\bar \lambda}_{+,i}\lambda_{+,i}-{\bar \lambda}_{-,i}\lambda_{-,i}\right)$. Surprisingly, the spin operator $S_{z}=\sum_{i}S_{z,i}$ is half of the generator of axial R-symmetry. Thus, for $U(k)$ gauge theories or ${\rm A}_{1}$ quiver gauge theories, we claim that:
\begin{equation*}
 \textit{ The symmetry of the Hamiltonian corresponds to the axial R-symmetry of gauge theory.}
\end{equation*}
 \noindent For instance, the Hamiltonian of the ${\rm A}_{1}$ XXX spin chain has a $SU(2)$ symmetry. It suggests that the low energies target space of gauge theory should be a hyperK\"{a}hler manifold such that the $U(1)$ axial R-symmetry will become a $SU(2)$ enhanced one at the conformal symmetry point. While the Hamiltonian of XXZ model has a $U(1)$ symmetry, which is exactly the axial $U(1)$ R-symmetry of gauge theory. Finally, we want to emphasize again that the axial R-symmetry relies on the quotient of the isometry of $C$ by its translation symmetry which is responsible for the Kaluza-Klein modes.

 For a general quiver gauge theory, we \textit{conjecture} that the spin operators could also depend on one more index: $\lambda^{n}_{\pm}$ and $\bar\lambda^{n}_{\pm}$, where $n$ and $\pm$ together label a finite subgroup of $SU(2)$ associated with the $ADE$ classification of a quiver diagram. Although a complete study of the two-dimensional quiver gauge theory is still missing, we expect that its low-energy SCFT, an emergent XXX spin chain, has an $ADE$ group as \textit{global symmetry}. The three-dimensional lifting, XYZ spin chain, has an affine $ADE$ group, while the four-dimensional lifting has the elliptic one. Our clue of this expectation comes from the study of Seiberg-Witten quiver gauge theories in \cite{Nekrasov:2012xe} and their corresponding integrable systems \cite{ Nekrasov:2013xda}.

 The finite symmetries ${\cal C}$, ${\cal P}$, and ${\cal T}$ descending from a quantum field theory can also be used to determine the Hamiltonian of an emergent spin chain. We will see that the time reversal, ${\cal T}$, could be broken in the case with a twisted boundary condition of a spin chain\footnote{It is also referred to as an open spin chain.}. In this situation, the Hamiltonian is not necessarily to be a hermitian one, namely a complex operator or its hermitian conjugate. While ${\cal P}$ ``symmetry" is extremely interesting. For example, if we operate this symmetry to gauge theory, it maps one gauge theory to its Seiberg(-like) dual one. However, we do not usually view the duality between two different quantum field theories as a symmetry. But from the perspective of the spin chain, it is a symmetry in the closed spin chain! This observation encourages us to take a further step. We conjecture \textit{those gauge theories with different ranks of the gauge group, but an identical global symmetry can be unified into a single framework: the Heisenberg spin chain}.

 Finally, we want to mention that Nekrasov and Shatashvili \cite{Nekrasov:2009uh, Nekrasov:2009ui, Nekrasov:2009rc} have uncovered several similarities between gauge theories and integrable systems before us. They mainly focus on the gauge theories flow to 2d SCFTs in the far infrared. Their main observations can be summarized in the following table:
   \begin{equation}\label{}
   \begin{tabular}{|ccc|}
   \hline
    ${\rm Y}(p_{a})$ & $\leftrightarrow$ &$\widetilde{ W}_{{\rm eff}}(\sigma_{a})$ \\
     $p_{a}$ &  $\leftrightarrow$ & $\sigma_{a}$\\
     $k$-{\rm particle sector} &  $\leftrightarrow$ & {\rm gauge group} $U(k)$\\
       $N$-{\rm sites} &  $\leftrightarrow$ & {\rm flavor group} $SU(N)$\\
       {\rm twisted boundary} &  $\leftrightarrow$ & $t=r-i\theta$\\
        {\rm (in-)homogeneities} &  $\leftrightarrow$ & {\rm twisted masses}\\
   \hline
 \end{tabular}
 \end{equation}
  where $p_{a}$ denote the rapidities, ${\rm Y}(p_{a})$ is the Yang-Yang function, and $\widetilde{ W}_{{\rm eff}}(\sigma_{a})$ is the twisted effective superpotential on the Coulomb branch. Thus, they claim that the vacuum equations
  \begin{equation}\label{}
    \exp\left(\frac{\partial\widetilde{ W}_{{\rm eff}}(\sigma_{a})}{\partial\sigma_{a}}\right)=1
  \end{equation}
  are exactly Bethe ansatz equations. On the other hand, our paper provides a framework guided by the global symmetries and the dynamics constraints of domain walls which serves as another physical explanation of observations found by Nekrasov and Shatashvili. One difference is that we keep and use fermionic degrees of freedom descending from field theory to describe a Heisenberg spin chain. So the appearance of a spin chain from gauge theory is \textit{more manifest} in our route. By contrast, the authors in previous studies\cite{Nekrasov:2009uh, Nekrasov:2009ui, Nekrasov:2009rc, Nekrasov:2010ka} preferred to use the bosonic fields solely to connect with an integrable system.

  The organization of this paper is as follows: In section \ref{Sec:2}, we not only review some basics of supersymmetric gauge theory with four supercharges but also give new results. They include BPS domain walls of nonabelian gauge theory, the interacting Hilbert space of gauge theory at an intermediate scale, etc. These new results serve the following sections. Section \ref{Sec:3} is devoted to presenting a framework to construct the Heisenberg spin chain from a supersymmetric gauge theory with four supercharges. Furthermore, in section \ref{Sec:4}, we state that the infrared duality between two gauge theories can be regarded as a symmetry in the closed spin chain. Then we further conjecture those gauge theories, which have the same kinds of representations but different ranks of gauge groups, are unified into a single framework: the Heisenberg spin chain. Finally, in the last two sections, we try to connect with other formulations of an integrable system. We provide new insight into the Yang-Baxter equation in section \ref{Sec:5} and postulate a possible path from our framework to the four-dimensional Chern-Simons theory in section \ref{Sec:6}.

\section{Supersymmetric Gauge Theories With Four Supercharges}\label{Sec:2}
This section is devoted to describing some background of supersymmetric gauge theories with four supercharges. However, some of the results introduced in this section are new. They may be useful for readers to understand the several new observations found in this paper.

In section \ref{Sec:2.1.1}, we will introduce some basics of gauged linear sigma models (GLSMs). Following the knowing result of solitons in abelian gauge theories, we will define domain walls in nonabelian gauge theories. Then we express the ground state wave function of supersymmetric gauge theories at the intermediate scale where the left fundamental degrees of freedom are domain walls. And in the following subsections, we will mainly focus on ${\cal N}$=(2,2) gauge theory from higher dimensions. For example, section \ref{Sec:2.2} has several aspects of 3d ${\cal N}$=2 Chern-Simons matter theories which relate to our paper. While section \ref{Sec:2.3} is for 4d ${\cal N}$=1 gauge theories. Finally, we summarize some other ways to reduce a higher-dimensional supersymmetric gauge theory to the two-dimensional ${\cal N}$=(2,2) one in section \ref{Sec:2.4}.

\subsection{Gauged Linear Sigma Models}\label{Sec:2.1}
GLSMs \cite{Witten:1993yc} were first proposed by Witten, who wanted to prove the correspondence between nonlinear sigma models and Landau-Ginzburg models conjectured in \cite{Greene:1988ut}. They then have been extensively studied in the context of nonabelian gauge theories \cite{Witten:1993xi,Hori:2006dk,Donagi:2007hi,Hori:2011pd,Jockers:2012zr}. Moreover, one can derive mirror symmetry \cite{Hori:2000kt,Gu:2018fpm} and compute Gromov-Witten invariants \cite{Jockers:2012dk,Gu:2020ana} by using GLSMs. However, we will not give a complete review in this section. Instead, we will focus on some unknown aspects of GLSMs that are useful for understanding the link between gauge theories and integrable systems.

In two-dimensional supersymmetric theory, ${\cal N}=(2,2)$ supersymmetry is generated by the fermionic charges $Q_{\pm}$, ${\bar Q}_{\pm}$. Classically, we also have vector and axial R-symmetries: $F_{V}$ and $F_{A}$. Those Noether charges satisfy the following relations:

\begin{eqnarray}
\{Q_{\pm}, {\bar Q}_{\pm} \} &=&\left(H \pm P\right), \\\nonumber
  \{{\bar Q}_{+}, {\bar Q}_{-} \}=Z, \quad&& \{Q_{+}, Q_{-}\}=Z^{*}, \\\nonumber
   \{{\bar Q}_{+},  Q_{-} \}=\widetilde{Z}, \quad&& \{ {Q}_{+}, {\bar Q}_{-} \}=\widetilde{Z}^{*},
\end{eqnarray}
and
\begin{eqnarray}
   [F_{V}, Q_{\pm}]=- Q_{\pm}, \quad&& [F_{V}, {\bar Q}_{\pm}]={\bar Q}_{\pm}  \\\nonumber
   [F_{A}, Q_{\pm}]=\mp Q_{\pm}, \quad&& [F_{A}, {\bar Q}_{\pm}]=\pm{\bar Q}_{\pm}.
\end{eqnarray}
The notation of $\pm$ in supercharges indicates their spins under a Lorentz transformation. In contrast with previous studies, the central charge $Z$ or $\widetilde{Z}$ will play a crucial role in understanding the integrability. More can be found in the book\cite{Hori:2003ic}. One well-known fact is that $F_{A}$ could be anomalous in a quantum theory. However, another fun fact is also a key to constructing the integrable system from a higher-dimensional gauge theory: In KK-reduction, the isometry of the extra dimensions relates to the axial R-symmetry or the vector R-symmetry in the mirror. \\

\noindent{\underline{Fields}}\\

We mainly focus on GLSMs with a $U(k)$ gauge group. However, the study for a general gauge group is similar. Matter fields in gauge theory are representations of the gauge group. And they are also chiral multiplets defined in the supersymmetric model. We denote them by $\Phi$. A chiral superfield can be expanded as
 \begin{equation}\label{}
   \Phi=\phi(z)+\theta^{+}\psi_{+}(z)+\theta^{-}\psi_{-}(z)+\theta^{+}\theta^{-}F(z),
 \end{equation}
 where
 \begin{equation}\label{}
   z^{\pm}=x^{\pm}-i\theta^{\pm}{\bar \theta}^{\pm}.
 \end{equation}

 The gauge field $A_{\mu}$ is a component of the vector multiplet denoted by $\mathrm{V}$, and its associated super gauge field strength are twisted chiral multiplet $\Sigma=\frac{1}{2}\{{\bar{\cal D}}_{+}, {\cal D}_{-}\}$ that can be written in terms of components:

\begin{equation}\label{}
   \Sigma=\sigma(\widetilde{z})+i\theta^{+}{\bar \lambda}_{+}(\widetilde{z})-i{\bar\theta}^{-}\lambda_{-}(\widetilde{z})+\theta^{+}{\bar\theta}^{-}\left(D(\widetilde{z})-iF_{01}(\widetilde{z})\right),
 \end{equation}
where
\begin{equation}\label{}
  \widetilde{z}^{\pm}=x^{\pm}\mp i\theta^{\pm}{\bar \theta}^{\pm},
\end{equation}
and
\begin{equation}\label{}
  F_{01}=\partial_{0}A_{1}-\partial_{1}A_{0}+[A_{0},A_{1}].
\end{equation}
\\
\
\\

\noindent{\underline{Lagrangian}}\\

  The Lagrangian of the corresponding two-dimensional quantum field theory has the kinetic-terms
  \begin{equation}\label{}
    \int d^{4}\theta {\rm Tr}\left({\bar\Phi} e^{V}\Phi\right)-\frac{1}{2e^{2}}{\rm Tr}{\bar\Sigma}\Sigma,
  \end{equation}
 the superpotential
  \begin{equation}\label{}
    \int d^{2}\theta W(\Phi)+c.c.,
  \end{equation}
  and the classical twisted superpotential
  \begin{equation}\label{}
    \int d^{2}\widetilde{\theta}\left(-\frac{t}{2}{\rm Tr}\Sigma \right)+c.c..
  \end{equation}
  The complex parameter $t$ is a linear combination of the FI parameter $r$ and the theta parameter $\theta$
  \begin{equation}\label{}
    t=r-i\theta.
  \end{equation}
  \\

\noindent{\underline{Flavor Symmetries And Twisted Masses}}\\

The field space may have a global symmetry $T$ that can be weakly gauged by turning on the background vector field $V_{T}$. Then it induces a twisted mass term in the superspace formulation
\begin{equation}\label{}
  \int d^{4}\theta {\rm Tr_{T}}{\bar \Phi}e^{V_{T}}\Phi,
\end{equation}
where
\begin{equation}\label{}
  V_{T}=\theta^{+}{\bar \theta}^{-}m_{T}+c.c.
\end{equation}
  To preserve ${\cal N}=(2,2)$ supersymmetry, the matrix $m_{T}$ should be diagonalizable with constant parameters:
  \begin{equation}\label{}
      m_{T}=\left(
    \begin{array}{ccc}
      m_{1} &  &  \\
       & \ddots &  \\
       &  & m_{{\rm rank\left(T\right)}} \\
    \end{array}
  \right).
  \end{equation}
 The overall shift of $m_{T}$ by the constant matrix $c\cdot\textbf{I}$ can be absorbed by a redefinition of the $\Sigma$ field, where $\textbf{I}$ is an identity matrix with the same rank as ${\rm rank\left(T\right)}$. These parameters are called twisted masses \cite{Alvarez-Gaume:1983uye, Hanany:1997vm}.\\

 \subsubsection{Abelian Gauge Theories}\label{Sec:2.1.1}
 We first focus on well-studied aspects of abelian gauge theories, and then we move to nonabelian gauge theories with new results presented.\\

 \noindent{\underline{Quantum Dynamics Of Abelian Gauge Theories}}\\

 The quantum behaves of the abelian gauged linear sigma model are also well-studied, and many exact results are known. We first start from abelian gauge theories that flow to mass-gap theories in the far infrared. For illustrative purposes, we first consider a $U(1)$ gauge theory with $N$ charge-one matters. Semiclassically, one can read off the vacuum structures from the potential energy
 \begin{equation}\label{PE}
   U=\frac{e_{{\rm eff}}^{2}}{2}\left(\sum^{N}_{i=1}\mid \phi_{i}\mid^{2}-r\right)^{2}+\sum^{N}_{i=1}\mid\sigma+m_{i}\mid^{2}\mid\phi_{i}\mid^{2},
 \end{equation}
 where the FI parameter $r$ is perturbative renormalized $ t(\mu)=r-i\theta=N\log\frac{\mu}{\Lambda}$. A complexified dynamical scale $\Lambda$ has been introduced. It is an RG-invariant quantity that can be defined either by theory at the cutoff or the physical scale
 \begin{equation}\label{}
    \Lambda=\mu e^{-t(\mu)}.
 \end{equation}
If the twisted mass $m_{N}\gg m_{i\neq N}$, the field $\phi_{N}$ will be dynamically frozen. Now, the mass $m_{N}$ is a new cutoff of the theory, and another RG-invariant scale emerges
\begin{equation}\label{}
  \Lambda^{\prime}=m_{N}e^{-t^{\prime}(m_{N})},\quad {\rm where}\quad \Lambda^{N}=m_{N}\left(\Lambda^{\prime}\right)^{N-1}.
\end{equation}
However, it was observed in \cite{Gu:2021beo} that the field strength $F_{01}$ is also a (pseudo)-scalar that can be used to parameterize the vacuum configuration. Since the coupling between the gauge field and the phase $\varphi$ of the matter field $\phi=\rho e^{i\varphi}$, it would be easier to work with the dual variable $\vartheta$ which we refer to as the dynamical theta angle. See \cite{Hori:2000kt} for more. Then, the modified vacuum potential energy is
 \begin{equation}\label{MPE}
   U=\frac{e_{{\rm eff}}^{2}}{2}\left[\left(\sum^{N}_{i=1}\mid \phi_{i}\mid^{2}-r\right)^{2}+\left(\sum^{N}_{i=1}\vartheta_{i}-\theta-2n\pi\right)^{2}\right]+\sum^{N}_{i=1}\mid\sigma+m_{i}\mid^{2}\mid\phi_{i}\mid^{2}.
 \end{equation}
 The perturbation theory is well-defined in UV where $e_{\rm eff}\sqrt{r}\gg\mu\gg\Lambda$. For generic twisted masses $m_{i}\neq m_{j}$, we have $N$ isolated vacua labeled by
 \begin{equation}\label{}
   \sigma=-m_{i},\quad\quad \mid\phi_{i}\mid^{2}=r\delta_{ij}.
\end{equation}
If the twisted masses are vanishing, one can find a continuous vacuum that is the projective space $\mathbb{P}^{N-1}$. However, if we include the full non-perturbative quantum correction, the vacua will be isolated even in UV. To see this, we first notice that one can group the field variables into the holomorphic twisted chiral fields $Y_{i}$ with the lowest components to be
 \begin{equation}\label{}
   y_{i}=\mid \phi_{i}\mid^{2}-i\vartheta_{i}.
 \end{equation}
 Then following Seiberg's idea of obtaining the exact results in 4d ${\cal N}=1$ supersymmetric gauge theories \cite{Seiberg:1994bz}, one eventually finds a two-dimensional Landau-Ginzburg theory that captures the whole quantum theory of the original gauged linear sigma model. It is defined on the target space $\left(\mathbb{C}^{\ast}\right)^{N}\times\mathbb{C}$ with the twisted superpotential
 \begin{equation}\label{TS}
   \widetilde{W}\left(\Sigma,Y_{i}; t\right)=\Sigma\left(\sum^{N}_{i=1}Y_{i}-t\right)+\sum^{N}_{i=1}\mu e^{-Y_{i}}+\sum^{N}_{i=1}m_{i}Y_{i}.
 \end{equation}
By mapping the twisted chiral fields $Y_{i}$ and $\Sigma$ to chiral ones, one can then define the 2d mirror Landau-Ginzburg theory for GLSM for $\mathbb{P}^{N-1}$, which is exactly how Hori and Vafa performed in deriving the abelian mirror symmetry. A subtle fact in the mirror is that the manifest global symmetry is, instead of the entire group, the maximal torus of the flavor group $SU(N)$ plus its Weyl symmetry and center group \cite{Gaiotto:2015aoa}. However, one can show that spectra in the mirror are representations of $SU(N)$.

So far, we have discussed the exact quantum theory of a massive gauge theory at the scale $\mu\gg \Lambda$. However, no semi-classical supersymmetric vacuum configuration exists in the region $\mu\ll \Lambda$ from the potential energy. But it was observed by Witten that the $N$ emergent vacua reappear in the effective theory on the Coulomb branch by integrating out matter fields
\begin{equation}\label{ETC}
  \int d^{4}\theta K\left(\Sigma,{\bar\Sigma}\right)+\frac{1}{2}\int d^{2}\widetilde{\theta}\widetilde{W}\left(\Sigma; m_{i}\right)+c.c.,
\end{equation}
where the twisted superpotential $\widetilde{W}\left(\Sigma; m_{i}\right)$ can be calculated exactly:
\begin{equation}\label{}
  \widetilde{W}\left(\Sigma; m_{i}\right)=-t\Sigma-\sum^{N}_{i=1}\left(\Sigma+m_{i}\right)\left(\log\left(\frac{\Sigma+m_{i}}{\mu}\right)-1\right).
\end{equation}
On the other hand, if we start from the exact theory (\ref{TS}), we can see $N$ vacua at almost every scale by computing
\begin{equation}\label{}
  \exp\left(d \widetilde{W}\left(Y_{i};\Sigma\right)\right)=1.
\end{equation}
After integrating out $Y_{i}$ fields, it reduces to the effective theory (\ref{ETC}).\\

\noindent{\underline{Twisted Chiral Rings}}\\

Classically, we have infinite twisted chiral rings labeled by
\begin{equation}\label{}
  \sigma^{l},\quad\quad l\in \mathds{Z}^{*}.
\end{equation}
However, the vacuum equation
\begin{equation}\label{}
  e^{d\widetilde{W}}=1
\end{equation}
gives the twisted chiral ring relation
\begin{equation}\label{}
  \prod^{N}_{i=1}\left(\sigma+m_{i}\right)=\Lambda^{N},
\end{equation}
which suggests that the dimension of rings is finite in quantum theory.\\

\noindent{\underline{Domain Walls In $\mathbb{P}^{N-1}$ Model}}\\

Since we have isolated vacua in a massive theory, one expects there exist massive domain walls (solitons) that interpolate different vacua at the two spatial infinities $x^{1}\rightarrow \pm\infty$. Let us first restrict to the case with vanishing twisted masses. It was first discussed by Witten in \cite{Witten:1978bc} that the fundamental field $\Phi$ \footnote{This operator can be gauge-invariant by attaching a Wilson line to infinity.} is a BPS domain wall in the ${\cal N}=(2,2)$ theory, although $\phi$ is confined in the associated bosonic theory. This means the elementary fields $\phi^{i}$, $\psi^{i}$ constitute BPS doublets in the fundamental representation of the flavor symmetry $SU(N)$. To see these, we come back to the effective theory (\ref{ETC}), which has the vacua:
\begin{equation}\label{}
  \sigma=\Lambda e^{\frac{i\left(\theta+2\pi l\right)}{N}},\quad\quad l=0,\cdots,N-1.
\end{equation}
The equation of motion (\ref{ETC}) with respect to $A_{0}$ \cite{Witten:1978bc}:
\begin{equation}\label{EOMTF}
  \frac{\partial}{\partial x^{1}}\left(\frac{1}{e^{2}_{{\rm eff}}}F_{01}\right)+\frac{\partial \left(\theta-N\arg\left(\sigma\right)\right)}{\partial x^{1}}=0.
\end{equation}
Then, integrating (\ref{EOMTF}) over the spatial direction, and from Eq.(\ref{MPE}), we see that $F_{01}=e^{2}_{{\rm eff} }(\sum^{N}_{i=1}\vartheta_{i}-\theta-2n\pi)=0$ for the vacua at $x^{1}\rightarrow \pm\infty$. Thus, we find that $\arg\sigma(+\infty)-\arg\sigma(-\infty)=0$. It says the effective theory (\ref{ETC}) has no domain wall configuration.

However, if we include the ``integrating-out" massive matter fields to be the source term $j^{\mu}=i\phi^{\dagger}D^{\mu}\phi-i\left(D^{\mu}\phi\right)^{\dagger}\phi-{\bar \psi}\gamma^{\mu}\psi$ that minimally coupled to the $U(1)$ gauge $A_{\mu}$. Then the EOM with respect to $A_{0}$ has been modified to
\begin{equation}\label{EOMA}
  \frac{\partial}{\partial x^{1}}\left(\frac{1}{e^{2}_{{\rm eff}}}F_{01}\right)+\frac{\partial \left(\theta-N\arg\left(\sigma\right)\right)}{\partial x^{1}}+j^{0}=0.
\end{equation}
Thus, we have a more interesting equation:
\begin{equation}\label{FDE}
  \arg\sigma(+\infty)-\arg\sigma(-\infty)=\frac{2\pi}{N}\int dx^{1}j^{0}.
\end{equation}
From the above, one can observe that the field $\Phi$ interpolates the neighboring vacua $\sigma\left(-\infty\right)=\Lambda e^{\frac{2\pi i n}{N}} \rightarrow \sigma\left(\infty\right)=\Lambda e^{\frac{2\pi i(n+1)}{N}} $, while the field ${\bar\Phi}$ reverses the direction, and we call it the anti-domain wall in this paper. Likewise, domain walls interpolating vacua by $\ell $ steps, $\sigma\left(-\infty\right)=\Lambda e^{\frac{2\pi i n}{N}} \rightarrow \sigma\left(\infty\right)=\Lambda e^{\frac{2\pi i(n+\ell)}{N}}$, carry electronic charge $\ell$. It is known that the corresponding domain walls consist of BPS-saturated bound states of $\ell$ elementary fields, which transform as the $\ell$-th anti-symmetric representation of $SU(N)$. The masses of these domain walls, up to an overall normalization, have been computed in several different contexts, see \cite{Koberle:1980wr, Abdalla:1983ae} and \cite{Cecotti:1991vb}, and they are
\begin{equation}\label{}
  \widetilde{Z}_{\ell_{1} \ell_{2}}=N\left|\Lambda\right|\left|e^{\frac{2\pi i(\ell_{1}-\ell_{2})}{N}}-1\right|.
\end{equation}\\
Finally, we comment that the two-dimensional Landau-Ginzburg theory (\ref{TS}) does have the field configurations corresponding to the domain walls, although they are not fundamental excitations. It is not surprising as the Laudau-Ginzburg model is an exact theory. See \cite{Hori:2000kt} for more details about the field configuration of domain walls in the mirror.\\

\noindent{\underline{BPS Spectrum In The $\mathbb{P}^{N-1}$ Model With Twisted Masses}}\\

If we include the twisted masses in gauge theory, the BPS spectrum has been computed in \cite{Hanany:1997vm} with a modified central charge
\begin{equation}\label{BPS}
  \widetilde{Z}_{\ell 0}=\Delta\widetilde{W}_{{\rm eff}}+2\pi i\sum^{N}_{i=1}m_{i}S_{i},
\end{equation}
where $S_{i}$ is the Noether charge of the $i$-th $U(1)$ of the group $U(1)^{N}$, the maximal unbroken torus group of the symmetry $U(N)$. See \cite{Dorey:1998yh} for more about the BPS spectrum (\ref{BPS}) and how it is similar to the BPS spectrum of Seiberg-Witten theory\cite{Seiberg:1994rs}.\\

\noindent{\underline{Comments On Vacua And Symmetries}}\\

We first consider the theory of vanishing twisted masses. The classical symmetry is $SU(N)\times U(1)_{A}\times U(1)_{V}$. The quantum correction suggests that $U(1)_{A}$ is anomalously broken to $\mathbb{Z}_{2N}$. In the far infrared, $\mu\rightarrow 0$, the theory will localize at a specific vacuum
\begin{equation}\label{}
  \sigma=\Lambda e^{\frac{2\pi i\ell}{N}}.
\end{equation}
Because $\sigma$ is charged under the left axial symmetry $\mathbb{Z}_{2N}$, the nonzero expectation value of $\sigma$ indicates a spontaneous breaking of the axial symmetry to $\mathbb{Z}_{2}$:
\begin{equation}\label{}
  \mathbb{Z}_{2N}\rightarrow\mathbb{Z}_{2},
\end{equation}
 which was emphasized before in the literature. However, we want to mention that since domain walls are infinite heavy in the far infrared, which suggests that we also have the other symmetry breaking:
\begin{equation}\label{}
  SU(N)\rightarrow \frac{SU(N)}{\mathbb{Z}_{N}}.
\end{equation}
In other words, the dynamics of domain walls, charged under the center group $\mathbb{Z}_{N}$, are frozen and can be regarded as probes of the theory. The order of the center group suggests that it has $N$ isolated vacua. Namely, the group acts on them transitively. If we turn on finite generic twisted masses, the $\mathbb{Z}_{N}$ symmetry will become a generic $N$-order group. So we still have $N$ vacua. This center symmetry is crucial for us to understand the domain walls of GLSMs that flow to SCFTs in the infrared.  \\

\noindent{\underline{GLSM For Tot ${\cal O}(-1)^{N}$}}\\

GLSM could flow to an SCFT in the infrared. For example, consider the GLSM for Tot ${\cal O}(-1)^{N}$, which is a $U(1)$ gauge theory with $N$ positive charge-one matters $\Phi_{i}$ and $N$ negative charge-one matters $\widetilde{\Phi}_{j}$. The Higgs vacuum configuration is an SCFT with the target space as the non-compact space Tot ${\cal O}(-1)^{N}$. The global symmetry is $SU(N)\times SU(N) \times U(1)_{a}\times U(1)_{A}\times U(1)_{V}$. The factor $SU(N)\times SU(N) \times U(1)_{a}$ is the semi-simple part of the flavor symmetry $U(N)\times U(N)$ with the vector combination $U(1)$-factor has been Higgsed by the $U(1)$ gauge group. A generic twisted mass breaks $SU(N)\times SU(N)$ to the maximal torus, we can then have isolated vacuum solutions on the Coulomb branch. To see this, after integrating out matters, it reduces to an effective theory on the Coulomb branch with the twisted effective superpotential
\begin{equation}\label{}
  \widetilde{W}_{{\rm eff}}=-t\Sigma-\sum^{N}_{i=1}\left(\Sigma+m_{i}\right)\left(\log\left(\Sigma+m_{i}\right)-1\right)-\sum^{N}_{j=1}\left(-\Sigma+\widetilde{m}_{j}\right)\left(\log\left(-\Sigma+\widetilde{m}_{j}\right)-1\right),
\end{equation}
where $m_{i}$ and $\widetilde{m}_{i}$ are twisted masses of $N$ positive charge-one matters and $N$ negative charge-one matters, respectively. The vacuum equation is
\begin{equation}\label{VETP}
  \prod^{N}_{i=1}\left(\frac{\sigma+m_{i}}{-\sigma+\widetilde{m}_{i}}\right)=q=e^{-t}.
\end{equation}
However, to integrate out matter fields, it is not necessary to require twisted masses that are all different. For example, one can set $\widetilde{m}_{i}=m_{i}=m$, then (\ref{VETP}) becomes
\begin{equation}\label{}
  \left(\frac{\sigma+m}{-\sigma+m}\right)^{N}=q.
\end{equation}
In this condition, we have a $\mathbb{Z}_{N}$ center symmetry. The equation of motion with respect to $A_{0}$ field is
\begin{equation}\label{EOMA}
  \frac{\partial}{\partial x^{1}}\left(\frac{1}{e^{2}_{{\rm eff}}}F_{01}\right)+\frac{\partial \left(\theta-N\arg\left(\sigma+m\right)+N\arg\left(-\sigma+m\right)\right)}{\partial x^{1}}+j^{0}=0.
\end{equation}
Now, integrating over $x^{1}$, we have
\begin{equation}\label{FDE}
  \arg\left(\left(\sigma+m\right)\left(-\sigma+m\right)^{-1}\right)(+\infty)- \arg\left(\left(\sigma+m\right)\left(-\sigma+m\right)^{-1}\right)(-\infty)=\frac{2\pi}{N}\int dx^{1}j^{0}.
\end{equation}
Thus, we find the domain wall $\Phi$ or $\bar{\widetilde{\Phi}}$ can both interpolate two adjacent vacua
\begin{equation}\label{}
  \frac{\left(\sigma+m\right)}{\left(-\sigma+m\right)}(-\infty)=q^{\frac{1}{N}}e^{\frac{2\pi i n}{N}}\rightarrow  \frac{\left(\sigma+m\right)}{\left(-\sigma+m\right)}(+\infty)=q^{\frac{1}{N}}e^{\frac{2\pi i (n+1)}{N}}.
\end{equation}
If $m$ has the same scaling behavior as $\sigma$, the central charges of the domain walls are vanishing because the theory flows to SCFT in the low energies. A crucial difference to the mass-gap theory is that we can not localize an SCFT to a specific vacuum in the far infrared, and $U(1)_{A}$ is not broken. However, we can still use the center symmetry $\mathbb{Z}_{N}$ to label $N$ different vacua in this situation.\\

\subsubsection{Nonabelian Gauge Theories}\label{Sec:2.1.2}
The research on nonabelian gauge theories is still active\cite{Gu:2020oeb}. For our aims, we first consider a well-studied example: GLSM for Grassmannian $Gr(k;N)$. It is a $U(k)$ gauge group with $N$ fundamental fields $\Phi_{i}$. Like in the abelian case, we first need to understand the vacuum structure of a quantum field theory. The semiclassical potential energy is
\begin{equation}\label{PNGT}
  U=\frac{e^{2}_{{\rm eff}}}{2} {\rm Tr}\left(\sum^{N}_{i=1}{\bar\phi}_{i}\phi_{i}-r\right)^{2}+\frac{1}{2e^{2}_{{\rm eff}}}{\rm Tr}\left[\sigma, {\bar\sigma}\right]^{2}+\sum^{N}_{i=1}{\bar\phi}_{i}\{\sigma, {\bar\sigma}\}\phi_{i}.
\end{equation}
The vanishing of the second term says that $\sigma$ must be diagonalizable:
\begin{equation}\label{VES}
  \sigma=\left(
    \begin{array}{ccc}
      \sigma_{1} &  &  \\
       & \ddots &  \\
       &  & \sigma_{k} \\
    \end{array}
  \right).
\end{equation}
 Meanwhile, the vanishing of the third term says: $\sigma$ is vanishing if some $\phi$ is nonzero. When $r\gg 0$ and $N\geq k$, it has a continuous vacuum configuration: the Grassmannian. As we discussed in the abelian gauge theory, if we include the full non-perturbative correction, the vacua are isolated, and the expectation value of $\sigma$ would be nonzero. Before seeing the exact result, we first consider the effective theory at the LG-point: $r\ll 0$. In this region, one can integrate out matters since they are heavy. The gauge field has two parts: one is the diagonal part, and the other one is the off-diagonal part, namely the W-bosons in this paper. Since the constraint (\ref{VES}) should be satisfied for any $r$, one may expect that the semiclassical expectation values of W-bosons are vanishing. Furthermore, we want to argue the vacua locate in the field space where $\sigma_{a}\neq\sigma_{b}$ if $a\neq b$. To see this, we first consider a locus where $\sigma_{a}=\sigma_{b}$ for some $a\neq b$, then besides the massive vacua, the effective theory could also have two massless W-bosons: $W_{ab}$ and $W_{ba}$. However, this can not happen. Because if we have massless fields in the far infrared, then it contradicts the theory at $r\gg 0$, where it is a compact theory. However, we do not expect any phase transition when $r$ varies. Thus, at $r\ll 0$, it is a massive effective theory on the Coulomb branch $\left(\mathbb{C}^{\ast}\right)^{k}/S_{k}$ with the twisted effective superpotential and excluded locus:
\begin{equation}\label{TES}
  \widetilde{W}_{{\rm eff}}\left(\Sigma_{a}; t\right)=-\left(t+i\pi(k-1)\right)\sum^{k}_{a=1}\Sigma_{a}-\sum^{k}_{a=1}N\Sigma_{a}\left(\log\frac{\Sigma_{a}}{\mu}-1\right),\quad\quad \Sigma_{a}\neq\Sigma_{b} \quad \rm{if} \quad a\neq b.
\end{equation}
The vacuum equations are
\begin{equation}\label{VEGR}
  \left(\sigma_{a}\right)^{N}=(-1)^{k-1}\Lambda^{N}, \quad\quad a\in\{1,\cdots,k\}.
\end{equation}
The solution number is $N$ choose $k$, which is exactly the Witten index of Grassmannian. On the other hand, the semi-classical low energy theory, at $r\gg 0$, is a nonlinear sigma model on Grassmannian.

The next question is whether we have an exact theory for a nonabelian gauge theory. It has been solved in \cite{Gu:2018fpm} in the context of mirror symmetry which is defined on the target space $\left(\mathbb{C^{\ast}}\right)^{kN}/S_{k}\times \mathbb{C}^{k}/S_{k}\times \mathbb{C}^{k(k-1)}/S_{k}$ with the twisted superpotential
\begin{eqnarray}\label{TSGR}
  \widetilde{W}\left(\Sigma_{a}, Y_{ib}, W_{cd};t\right) &=& \sum^{k}_{a=1}\Sigma_{a}\left(\sum_{i,b}\rho^{a}_{ib}Y_{ib}+\sum_{b\neq c}\alpha^{a}_{bc}\ln W_{bc}-t\right) \\\nonumber
   &+&  \mu\sum_{a\neq b} W_{ab}+\mu\sum_{i,a}e^{-Y_{ia}},
\end{eqnarray}
where $\rho^{a}_{ib}=\delta^{a}_{b}$ are the weights of the fundamental representation of the gauge group $U(k)$, and $\alpha^{a}_{bc}=-\delta^{a}_{b}+\delta^{a}_{c}$ are the roots of the gauge group $U(k)$. The Weyl group $S_{k}$ acts on the fields as
\begin{equation}\label{}
  \Sigma_{a}\mapsto\Sigma_{b},\quad\quad Y_{ia}\mapsto Y_{ib},\quad\quad \sum_{c\neq d}\alpha^{a}_{cd} W_{cd}\mapsto \sum_{c\neq d}\alpha^{b}_{cd} W_{cd}
\end{equation}
This theory can be defined at almost every scale except the extreme UV, where the gauge coupling $e^{2}$ goes to 0, and the vacuum equations are
\begin{equation}\label{VEGRF}
  e^{-y_{ia}}=x_{a}=\sigma_{a}, \quad\quad w_{ab}=\sigma_{a}-\sigma_{b},\quad\quad \left(x_{a}\right)^{N}=(-1)^{k-1}\Lambda^{N}.
\end{equation}
From the twisted superpotential (\ref{TSGR}), the locus, $W_{ab}=0$, is dynamically excluded. This means the vacua only locate at
$\sigma_{a}\neq\sigma_{b}$, so Eq.(\ref{VEGRF}) has the same vacuum equations as in (\ref{VEGR}). The study of GLSM for a general target is similar.

 Before to new results, we want to give several comments:\\

$\bullet$ The abelian mirror of the GLSM for a toric variety is a fundamental theory \footnote{The mirror of GLSM with a superpotential is an effective theory in the UV, but it has a run-away vacuum configuration at a singular locus. However, since it flows to an SCFT in the far infrared, the field variables in the IR can touch that locus \cite{Morrison:1995yh}. This can also be understood in the semi-classical analysis of gauge theory at the LG point, where a finite orbifold emerges in the Higgs mechanism. See \cite{Aharony:2016jki,Gu:2020ivl} for a similar phenomenon in the pure gauge theories.}, this means taking the extremal UV-limit $\mu\rightarrow\infty$, the theory is a free theory with no singular locus. While the full quantum theory of a nonabelian gauge theory is an effective theory since we have the dynamically excluded locus on the field space $W^{ab}$ and $\Sigma^{a}$. This indicates that the nonabelian mirrors miss degrees of freedom, although their absence does not affect the vacuum structure and BPS spectrum.

$\bullet$ Nonabelian mirrors as effective theories can be understood from the original idea in \cite{Gu:2018fpm}. In the study of nonabelian gauge theory, we first go to the ``generic Coulomb branch", where $\Sigma_{a}\neq \Sigma_{b}$ for $a\neq b$, then one can find masses of the diagonal part of the gauge field are different from the ones of W-bosons. Moreover, W-bosons are charged under the left gauge group $U(1)^{k}$, and only a chiral superfield has this property. Thus, it reduces to an abelian-like theory. Its gauge group is $U(1)^{k}\rtimes S_{k}$, with two types of chiral superfields: the original matters with gauge charges as the weights, while the other type has gauge charges to be the roots of the $U(k)$ group but with vector R-charge 2. With further help from the exact results in abelian theories, one can obtain the nonabelian mirrors. They are definitely effective theories in Wilson's sense. The existence of nonabelian mirrors teaches us that the semi-classical excluded locus, $\Sigma_{a}\neq \Sigma_{b}$ for $a\neq b$, is also reliable in an exact theory. Therefore, no vacuum locates at the excluded locus. In conclusion, although the classical (fundamental) theory defined in the extremal UV is free, the quantum theory has to omit an excluded locus on field space.

$\bullet$ In previous research, the assignment of vector R-charges 2 to these W-bosons superfields can be understood from the Higgs mechanism \cite{Benini:2016qnm}, RG-flows \cite{Gu:2020ana}, and the exact results of gauge theories \cite{Halverson:2013eua}. Here we give a direct understanding from the bare action. We expand every field $\chi=\chi^{a}H_{a}+e\chi^{\alpha}E_{a}$ in the Lagrangian of the gauge field
\begin{equation}\label{}
  \int d^{4}\theta -\frac{1}{2e^{2}}{\rm Tr}{\bar\Sigma} \Sigma,
\end{equation}
where $e$ is the gauge coupling, $H_{a}$ and $E_{\alpha}$ are the Cartan-Weyl basis with $a$ runs over the Cartan subalgebra, and $\alpha$ are the roots of the gauge group. By using the Cartan algebra
 \begin{equation}\label{}
   \left[H_{a}, E_{\alpha}\right]=\alpha_{a}E_{\alpha},\quad\quad \left[H_{a}, H_{b}\right]=0,\quad\quad \left[E_{\alpha}, E_{-\alpha}\right]=\frac{2}{\left|\alpha\right|^{2}}\alpha_{a}H_{a}.
 \end{equation}
One will find that the Lagrangian reduces to the quadratic kinetic terms of the gauge field $A^{\alpha}_{z}$, gaugino $\lambda^{\alpha}$, and their conjugates. See \cite{Closset:2015rna}[appendix C.4] for a calculation in the topological A-twisted background. Of course, there are other quadratic terms and higher-order interactions in the Cartan-Weyl basis. However, the contributions of those quadratic terms to path integrals will be canceled by the gauge-fixing term. And the higher-order interactions can be discarded by supersymmetric localization. Furthermore, the charges of super-coordinates are
\begin{equation*}
  \begin{tabular}{|c|c|c|c|c|}
    \hline
      & $\theta^{-}$ & ${\bar\theta}^{+}$ & ${\bar\theta}^{-}$ & $\theta^{+}$ \\
      $F_{V}$ & 1 & -1 & -1 & 1 \\
      $F_{A}$ & -1 & -1 & 1 & 1 \\
      $S_{E}$ & $-\frac{1}{2}$ & $\frac{1}{2}$ & $-\frac{1}{2}$ & $\frac{1}{2}$ \\
    \hline
  \end{tabular},
\end{equation*}
where $S_{E}$ measures the Lorentz charge. Then, we can read off charges of fermions in a vector multiplet:
\begin{equation*}
  \begin{tabular}{|c|c|c|c|c|}
    \hline
      & ${\bar\lambda}_{-}$ & $\lambda_{+}$ & $\lambda_{-}$ & ${\bar\lambda}_{+}$ \\
      $F_{V}$ & -1 & 1 & 1 & -1 \\
      $F_{A}$ & -1 & -1 & 1 & 1 \\
      $S_{E}$ & $\frac{1}{2}$ & $-\frac{1}{2}$ & $\frac{1}{2}$ & $-\frac{1}{2}$ \\
    \hline
  \end{tabular}.
\end{equation*}
 These charges are exactly the same as of fermions in a chiral superfield with vector R-charge 2 and axial R-charge 0:
 \begin{equation}
    \begin{tabular}{|c|c|c|c|c|}
    \hline
      & ${\bar\psi}_{-}$ & $\psi_{+}$ & $\psi_{-}$ & ${\bar\psi}_{+}$ \\
      $F_{V}$ & -1 & 1 & 1 & -1 \\
      $F_{A}$ & -1 & -1 & 1 & 1 \\
      $S_{E}$ & $\frac{1}{2}$ & $-\frac{1}{2}$ & $\frac{1}{2}$ & $-\frac{1}{2}$ \\
    \hline
  \end{tabular}.
 \end{equation}
  So it is natural to expect that the quantum behavior of a W-boson is similar to a vector R-charge 2 chiral superfield associated with the identical gauge charge of that W-boson. However, there are two issues in this argument. The first issue is that the interaction between a W-boson and the maximal torus part of the gauge field is different from the one for a chiral superfield. It can be resolved by supersymmetric localization: higher-order interactions will be suppressed in the exact quantum theory. The second issue is that the bosonic field of a W-boson has a vanishing R-charge but a spin one. While the lowest component of a vector R-charge 2 chiral multiplet is a scalar. So they look different. However, one can argue that their contributions to the path integral are, in fact, the same. To see this, let us start from the A-twisted background first. In this background, the actual physical charge is $s+F_{V}/2$. So the case with $s=1$ and $F_{V}=0$ is the same as the one with $s=0$ and $F_{V}=2$ \cite{Closset:2015rna}. While in the physical background, we notice that the partition function $Z={}_{R}\langle {\bar 0}|0\rangle_{R}$ \cite{Cecotti:1991me} could be expressed as gluing an A-twisted hemisphere with the other hemisphere in the ${\rm \bar{A}}$-twisted background. By using the statement in the A(${\rm \bar{A}}$)-twisted background, we conclude that the quantum behavior of a spin one gauge field is also the same as a vector R-charge 2 scalar in a physical theory. In conclusion, a W-boson multiplet can be regarded as a chiral superfield with vector R-charge 2 in the exact quantum field theory, even though they look different in the (semi-)classical sense.\\
\

\noindent{\underline{BPS Spectrum}}\\

Unlike the BPS spectrum studied in abelian GLSMs from many aspects, the BPS spectrum of nonabelian GLSMs has not been investigated in the literature. However, since the exact theory of nonabelian GLSM is equivalent to the one for an abelian-like gauge theory, the BPS spectra of a nonabelian gauged linear sigma model can be similarly discussed. Let us first study the Grassmannian. Focus on the K\"{a}hler potential of the maximal torus part of the nonabelian gauge field in the effective theory:
\begin{equation}\label{}
  -\sum^{k}_{a=1}\frac{1}{2e^{2}_{{\rm eff}; a}}{\bar \Sigma}_{a}\Sigma_{a}.
\end{equation}
The EOM with respect to the gauge field $A^{a}_{0}$:
\begin{equation}\label{SENA}
  \frac{\partial}{\partial x^{1}}\left(\frac{1}{e^{2}_{{\rm eff;a}}}F_{01;a}\right)+\frac{\partial\left(\theta-N\arg\left(\sigma_{a}\right)\right)}{\partial x^{1}}+j^{0}_{a}=0, \quad a\in\{1,\cdots,k\}.
\end{equation}
After integrating over the spacial coordinate $x^{1}$, we have
\begin{equation}\label{NDMW}
  \arg\left(\sigma_{a}\right)\left(+\infty\right)-\arg\left(\sigma_{a}\right)\left(-\infty\right)=\frac{2\pi}{N}\int dx^{1}j^{0}_{a}.
\end{equation}
So the field $\Phi^{a}$ interpolates two nearest-neighbor vacua $\sigma_{a}\left(-\infty\right)=\Lambda(-1)^{\frac{k-1}{N}}e^{\frac{2\pi in}{N}}\rightarrow \sigma_{a}\left(+\infty\right)=\Lambda(-1)^{\frac{k-1}{N}}e^{\frac{2\pi i\left(n+1\right)}{N}}$. Thus, the domain wall, for each specific index $a$, looks the same as the one in an abelian theory. Nevertheless, there is one issue in Eq.(\ref{SENA}). We will see later that there are off-diagonal components of the K\"{a}hler metric in the effective theory:
\begin{equation}\label{}
  -\frac{1}{2e^{2}_{{\rm eff};a\neq b}}{\bar\Sigma}_{a}\Sigma_{b}.
\end{equation}
So the precise one for Eq.(\ref{SENA}), in fact, is
\begin{equation}\label{SENAT}
  \frac{\partial}{\partial x^{1}}\left(\frac{1}{e^{2}_{{\rm eff;a}}}F_{01;a}+\sum_{b}\frac{1}{e^{2}_{{\rm eff;a\neq b}}}F_{01;b}\right)+\frac{\partial\left(\theta-N\arg\left(\sigma_{a}\right)\right)}{\partial x^{1}}+j^{0}_{a}=0, \quad a\in\{1,\cdots,k\}.
\end{equation}
But this will not affect our Eq.(\ref{NDMW}).

However, we have more constraints in a nonabelian gauge theory: the Weyl gauge symmetry $S_{k}$, and the dynamically excluded locus:
\begin{equation}\label{}
  \sigma_{a}\neq\sigma_{b},\quad\quad {\rm for} \quad\quad a\neq b.
\end{equation}
We gauge the Weyl symmetry by imposing the following order
\begin{equation}\label{WGF}
  \sigma_{a}=\Lambda (-1)^{\frac{k-1}{N}}e^{\frac{2\pi i n_{a}}{N}},\quad\quad  n_{a}<n_{b},\quad\quad {\rm if} \quad\quad a<b
\end{equation}
to label the vacua. Moreover, a gauge-invariant variable can be defined as
\begin{equation}\label{GIFV}
  \left(\prod^{k}_{a=1}\sigma_{a}\right),
\end{equation}
then the field $\Phi^{a}$ is infinitely heavy if the vacua are labeled by $n_{a}+1=n_{a+1}$. Or, it is finite and interpolates the adjacent vacua
\begin{eqnarray}\label{}
  \left(\prod^{k}_{a=1}\sigma_{a}\right)\left(-\infty\right)&=&\Lambda^{k}(-1)^{\frac{k(k-1)}{N}}e^{\frac{2\pi i \sum^{k}_{a=1}n_{a}}{N}}\\\nonumber
  &\mapsto& \left(\prod^{k}_{a=1}\sigma_{a}\right)\left(+\infty\right)=\Lambda^{k}(-1)^{\frac{k(k-1)}{N}}e^{\frac{2\pi i \left(\sum^{k}_{a=1}n_{a}+1\right)}{N}}.
\end{eqnarray}
By direct counting, one can show the number of domain walls with charge-one is
\begin{equation}\label{DCS1}
  {N \choose k}.
\end{equation}
One subtlety in the above calculation is about the domain walls that move $n_{k}=N$ to $n_{k}=N+1$. In order to maintain the gauge imposed in ({\ref {WGF}}), we map $n_{k}=N+1$ to $n'_{1}=1$ by using the periodic structure of the vacua and the Weyl symmetry.

The domain walls $\Phi^{a}$ is the fundamental representation of the flavor group $SU(N)$.  Because of the extra index $a$, it is not surprising that the number of charge-one domain walls in $Gr(k;N)$ is more than the ones in a projective space $\mathbb{P}^{N-1}$. The number of charge-one domain walls of $\mathbb{P}^{N-1}$ is $N$ by setting $k=1$ in Eq.(\ref{DCS1}), which is exactly the number of matters. The central charge of these domain walls can be computed directly
\begin{equation}\label{}
  \widetilde{Z}_{\left(l+1\right)l}=N\left|\Lambda(-1)^{\frac{k-1}{N}}\right|\left|e^{\frac{2\pi i}{N}}-1\right|,
\end{equation}
where $l=\sum^{k}_{a=1}n_{a}$. The domain wall with the gauge charge $\ell$ can be studied similarly. Finally, if turn on the twisted masses in the target space $Gr(k;N)$, the central charges are
\begin{equation}\label{}
  \widetilde{Z}_{\left(l+\ell\right)l}=\triangle\widetilde{W}_{{\rm eff}}+2\pi i\sum^{N}_{i=1}S_{i}m_{i}.
\end{equation}
The above expression looks similar to the central charge in $\mathbb{P}^{N-1}$ described in Eq.(\ref{BPS}). These new results of domain walls in nonabelian gauge theories are already enough for our purposes in this paper, and we leave the complete investigation of their dynamics to future work.\\

\noindent{\underline{GLSMs For $T^{\ast}Gr(k;N)$}}\\

Our last example is about GLSM for $T^{\ast}Gr(k;N)$, and when $k$=1, it is $\textbf{{\rm Tot}}\ {\cal O}(-1)^{N}$. The gauge group is $U(k)$, and the matter content includes $N$ fundamental representation $\Phi_{i}$, $N$ anti-fundamental representation $\widetilde{\Phi}_{i}$, and one adjoint representation $X$. There is a superpotential in the GLSM for this target:
\begin{equation}\label{TGRS}
  W=\sum^{N}_{i=1}{\rm Tr}_{U(k)}\widetilde{\Phi}_{i}X^{2s}\Phi_{i},
\end{equation}
where $2s$ is a positive integer. The geometrical phase has been investigated in \cite{Nekrasov:2009ui}, and the readers can find more from it. While we mainly focus on the symmetry of the model. As we discussed before for $\textbf{{\rm Tot}}\ {\cal }O(-1)^{N}$, the symmetry is
\begin{equation}\label{}
  SU(N)\times SU(N)\times U(1)_{a}\times U(1)_{A}\times U(1)_{V}.
\end{equation}
When turning on the superpotential in the $k>1$ case, the above symmetry is explicitly broken to
 \begin{equation}\label{}
   SU(N)\times U(1)_{2s}\times U(1)_{A}\times U(1)_{V},
 \end{equation}
 where the group $U(1)_{2s}$ acts on the fields as
 \begin{equation}\label{}
   U(1)_{2s}: \left(\widetilde{\Phi}, X, \Phi\right)\mapsto \left(e^{-i2s\cdot u}\widetilde{\Phi}, e^{+i2u}X, e^{-i2s\cdot u}\Phi\right).
 \end{equation}
 If we further turn on the twisted masses for the flavor group $SU(N)$, it is broken to the maximal torus. As discussed in \cite{Nekrasov:2009ui}, the vacuum configuration in the geometrical phase suggests that $\Phi$ and $\widetilde{\Phi}$ can take nonzero expectation values to label the geometrical target $T^{\ast}Gr(k;N)$, while $X$=0. This suggests that $U(1)_{2s}$ is broken to the center group $\mathbb{Z}_{2s}$ at the geometrical vacua. However, we follow \cite{Nekrasov:2009ui} to reduce the theory to an effective theory on the Coulomb branch. To do this, we assign the complexified twisted masses $m^{{\rm f}}_{i}=m_{i}-2su$ to the fundamental matters, $m^{{\rm \bar f}}_{i}=-m_{i}-2su$ to the anti-fundamental, and $m^{{\rm adj}}=2u$ to the adjoint. Since these GLSMs flow to SCFTs in the IR, the one-loop correction to the FI parameter can at most a constant. By turning on the twisted masses, we have the twisted effective superpotential
 \begin{eqnarray}\nonumber
   \widetilde{W}_{{\rm eff}}\left(\Sigma_{a}; t\right) &=& -\left(t+i(k-1)\pi\right)\sum^{k}_{a=1}\Sigma_{a}-\sum^{k}_{a,b=1}\left(\Sigma_{a}-\Sigma_{b}+2u\right)\left(\log\left(\Sigma_{a}-\Sigma_{b}+2u\right)-1\right)\\\nonumber
    && \quad -\sum^{k}_{a=1}\sum^{N}_{i=1}\left(-\Sigma_{a}-m_{i}-2su\right)\left(\log\left(-\Sigma_{a}-m_{i}-2su\right)-1\right)\\\nonumber
    &&\quad\quad    -\sum^{k}_{a=1}\sum^{N}_{i=1}\left(\Sigma_{a}+m_{i}-2su\right)\left(\log\left(\Sigma_{a}+m_{i}-2su\right)-1\right).
 \end{eqnarray}
The vacuum equations are
\begin{equation}\label{}
  \prod^{N}_{i=1}\left(\frac{\sigma_{a}+m_{i}-2su}{\sigma_{a}+m_{i}+2su}\right)=e^{-t}(-1)^{N}\prod_{b\neq a}\frac{\sigma_{a}-\sigma_{b}-2u}{\sigma_{a}-\sigma_{b}+2u}.
\end{equation}
If $u$ is sufficiently large and the parameter $q=e^{-t}$ is generic, the above twisted effective superpotential is even well-defined for vanishing $m_{i}$. The Witten index of this has been counted in \cite{Shu:2022vpk,Jiang:2017phk}. In this situation, the vacuum equations reduce to
\begin{equation}\label{}
   \left(\frac{\sigma_{a}-2su}{\sigma_{a}+2su}\right)^{N}=e^{-t}(-1)^{N}\prod_{b\neq a}\frac{\sigma_{a}-\sigma_{b}-2u}{\sigma_{a}-\sigma_{b}+2u}.
\end{equation}

\subsubsection{Hilbert Space}\label{Sec:2.1.3}

In previous sections, we mainly focused on the field-theoretic language. It certainly has many advantages in many aspects, as we can see from the historical developments of quantum field theory.  However, we could also describe the quantum theory in terms of Hilbert space. A physical observable in a quantum field theory will be a Hermitian operator acting on the Hilbert space. But it has an intrinsic difficulty to use this framework directly in an interacting quantum field theory. Traditionally, the computation of a physical quantity, such as the scattering amplitude in a perturbative quantum field theory, is not straightforward. The procedure is as follows: to define free particles first where the interactions have been turned off, and then ``adiabatic" turning on the coupling constant, such that one could map the states of free particles to the ones in an interacting Hilbert space. This is so-called the ``LSZ" formula.

However, in two-dimensional quantum field theory, one may have a chance to describe the dynamics directly in an interacting Hilbert space. One piece of evidence of this is the well-known state/operator correspondence in the two-dimensional CFT. In this paper, we want to convince the readers that a similar thing can happen in a massive two-dimensional gauge theory at the intermediate scale. We emphasize that the connection between gauge theories and integrable systems will be transparent in this way.\\

\noindent{\underline{Grassmannian}}\\

We first describe the ground state wave functions in the effective theory (\ref{TES}) defined at the physical scale $\mu\ll \mid\sigma_{a}\mid$ with the approximate K\"{a}hler metric for the diagonal components
\begin{equation}\label{}
  \partial_{a}{\bar\partial}_{ a}K_{{\rm eff}}=\frac{1}{2e^{2}_{{\rm eff}; a}}=\frac{1}{2e^{2}}+\frac{N}{4\mid\sigma_{a}\mid^{2}}+\sum_{b\neq a}\frac{1}{2\mid\sigma_{a}-\sigma_{b}\mid^{2}},
\end{equation}
and the off-diagonal part
\begin{equation}\label{}
  \partial_{a}{\bar\partial}_{b}K_{{\rm eff}}=-\frac{1}{2\mid\sigma_{a}-\sigma_{b}\mid^{2}}.
\end{equation}
The K\"{a}hler metric in the large $\sigma$ for the case of $\mathbb{P}^{N-1}$ has been calculated in \cite{Hori:2003ic}[Chapter 15.5]. The computation of the K\"{a}hler metric for a nonabelian gauge theory on the generic Coulomb branch should be similar. It has one more ingredient: integrating out W-bosons will introduce a shift to the K\"{a}hler metric
\begin{equation}\label{}
  \sim \frac{1}{\mid\sigma_{a}-\sigma_{b}\mid^{2}},
\end{equation}
and this means the field space $\Sigma_{a}$ is defined on the locus
\begin{equation}\label{}
  \Sigma_{a}\neq 0,\quad\quad \Sigma_{a}\neq\Sigma_{b},\quad\quad {\rm for} \quad\quad a\neq b.
\end{equation}
We can certainly continue integrating out the high-frequency modes of $\Sigma$ multiplet fields, then K\"{a}hler potential will be further corrected, but the twisted effective superpotential will not be affected. Although it is difficult to know the actual quantum correction to the K\"{a}hler potential, however, since $e^{2}_{{\rm eff}}$ has the mass dimension two, we expect that
\begin{equation}\label{}
  e^{2}_{{\rm eff}}\rightarrow \infty, \quad\quad {\rm when}\quad\quad \mu\rightarrow 0.
\end{equation}
The actual formula of $e^{2}_{{\rm eff}}$ is not crucial for our purposes in this paper, although it would be interesting to write out the exact RG-flow of the K\"{a}hler potential. For later use, we make the K\"{a}hler metric diagonal by changing variables. By abuse of notation, the effective action can be expressed as
\begin{eqnarray}\nonumber
   \int_{\geq \frac{1}{\mu}} &d^{2}x&\frac{1}{2e^{2}_{{\rm eff};a}}\left(-\partial^{\nu}{\bar \sigma}_{ a}\partial_{\nu}\sigma_{a}+i{\bar\lambda_{-; a}}\left(\partial_{0}+\partial_{1}\right)\lambda_{-;a}+i{\bar\lambda_{+; a}}\left(\partial_{0}-\partial_{1}\right)\lambda_{+;a} \right)  \\\nonumber
   &&-\frac{e^{2}_{{\rm eff};a}}{2}\left|\partial_{ a}\widetilde{ W}_{{\rm eff}}\left( \sigma\right)\right|^{2}-\left(\frac{1}{2}\partial_{a}\partial_{ b}\widetilde{ W}_{{\rm eff}}\left( \sigma\right){\bar \lambda}_{+,a} \lambda_{-,b}+c.c.\right).
\end{eqnarray}
The effective theory is defined in the Euclidean momentum shell $0\leq p^{2}\leq \mu^{2}$, and the spectrum certainly depends on the spacial momentum. In order to get the ground state wave functions in a far-infrared limit, one can drop the spatial dependence of the fields \cite{Witten:1982df}. The theory then reduces to a quantum mechanic system:
\begin{eqnarray}\nonumber
    \int_{\geq \frac{1}{\mu}}  &dt &\frac{L}{2e^{2}_{{\rm eff};a}}\left(-\partial^{0}{\bar \sigma}_{ a}\partial_{0}\sigma_{a}+i{\bar\lambda_{-; a}}\partial_{0}\lambda_{-;a}+i{\bar\lambda_{+; a}}\partial_{0}\lambda_{+;a} \right)  \\\nonumber
   && -\frac{Le^{2}_{{\rm eff};a}}{2}\left|\partial_{ a}\widetilde{ W}_{{\rm eff}}\left( \sigma\right)\right|^{2}-L\left(\frac{1}{2}\partial_{a}\partial_{ b}\widetilde{ W}_{{\rm eff}}\left( \sigma\right){\bar \lambda}_{+,a} \lambda_{-,b}+c.c.\right),
\end{eqnarray}
where $L$  is the spatial length. Because we only consider the tiny time-variation modes, one can regard the coupling $e^{2}_{{\rm eff}}$ as a constant. Therefore, we can do the following rescaling of the fields
\begin{equation}\label{}
  \sigma_{a}\mapsto\sqrt{\frac{2e^{2}_{{\rm eff}}}{L}}\sigma_{a},\quad\quad \lambda_{\pm}\mapsto\sqrt{\frac{2e^{2}_{{\rm eff}}}{L}}\lambda_{\pm},\quad\quad {\bar\lambda}_{\pm}\mapsto  \sqrt{\frac{2e^{2}_{{\rm eff}}}{L}}{\bar\lambda}_{\pm},
\end{equation}
the Lagrangian further reduces to
\begin{eqnarray}\label{LGRM}
  \int &dt& \left(\sum^{k}_{a=1}\left|\dot{\sigma}_{a}\right|^{2}+i{\bar\lambda_{-; a}}\partial_{0}\lambda_{-;a}+i{\bar\lambda_{+; a}}\partial_{0}\lambda_{+;a} \right)  \\\nonumber
   && -\frac{\zeta^{2}}{4}\left|\partial_{ a}\widetilde{ W}_{{\rm eff}}\left( \sigma\right)\right|^{2}-\left(\frac{\zeta}{2}\partial_{a}\partial_{ b}\widetilde{ W}_{{\rm eff}}\left( \sigma\right){\bar \lambda}_{+,a} \lambda_{-,b}+c.c.\right),
\end{eqnarray}
where $\zeta=\sqrt{2Le^{2}_{{\rm eff};a}}$. When $\mu\rightarrow 0$, since the mass dimension of $e^{2}_{{\rm eff};a}$ is two, we have $\zeta\rightarrow\infty$ even though $L\rightarrow 0$ in this limit. The readers may notice that Eq.(\ref{LGRM}) is related to the Morse theory if we identify
\begin{equation}\label{MHF}
  h\left(\sigma\right)={\rm Re}\widetilde{W}_{\rm eff}\left(\sigma\right).
\end{equation}
 We first consider the limit where $\zeta\rightarrow\infty$, then the theory will be localized at a specific vacuum, i.e. the critical points $\langle\sigma\rangle_{a}$
\begin{equation}\label{}
  e^{\partial_{a}\widetilde{W}_{{\rm eff}}\left(\sigma\right)}=1.
\end{equation}
Then expand the field $\sigma_{a}$ around the vacua, and define the local coordinates $z_{a}=x_{a}+iy_{a}=\sigma_{a}-\langle\sigma\rangle_{a}$
\begin{equation}\label{}
  \widetilde{W}_{{\rm eff}}\left(\sigma\right)=\widetilde{W}_{{\rm eff}}\left(\langle\sigma\rangle_{a}\right)+\sum^{k}_{a=1}c_{a}(z_{a})^{2}+{\cal O}(z^{3}_{a}),
\end{equation}
where $c_{a}=(-1)^{\frac{N-k+1}{N}}e^{-\frac{2\pi in^{a}}{N}}N\Lambda^{-1}$. For the Morse function $h$, we have
\begin{equation}\label{}
  h={\rm Re}\widetilde{W}_{{\rm eff}}\left(\langle\sigma\rangle_{a}\right)+\sum^{k}_{a=1}{\rm Re}\left(c_{a}(z_{a})^{2}\right)+\cdots.
\end{equation}
After diagonalizing the variables $z_{a}$ to $z'_{a}=x'_{a}+iy'_{a}$, we have
\begin{equation}\label{}
  h={\rm Re}\widetilde{W}_{{\rm eff}}\left(\langle\sigma\rangle_{a}\right)-\sum^{k}_{a=1}\frac{N}{\left|\Lambda\right|}\left((x'_{a})^{2}-(y'_{a})^{2}\right)+\cdots.
\end{equation}
Since the Morse index is $k$, the number of negative eigenvalues of the Hessian of $h$, the ground state wave functions must have fermion number $k$ \cite{Witten:1982im}. Furthermore, because the effective theory is reduced from a two-dimensional gauge theory, the global symmetries $F_{V}$ and $F_{A}$ in the gauge theory can be defined in supersymmetric quantum mechanics as well. They are
\begin{equation}\label{}
 F_{V}=-\sum^{k}_{a=1}\left({\bar\lambda}_{-,a}\lambda_{-,a}+{\bar\lambda}_{+,a}\lambda_{+,a}\right),\quad\quad F_{A}=\sum^{k}_{a=1}\left({\bar\lambda}_{+,a}\lambda_{+,a}-{\bar\lambda}_{-,a}\lambda_{-,a}\right)
\end{equation}
The R-charges of each fermionic field are
\begin{center}
  \begin{tabular}{|c|c|c|c|c|}
    \hline
      & ${\bar\lambda}_{+}$ & $\lambda_{-}$ & $\lambda_{+}$ & ${\bar\lambda}_{-}$ \\
    $F_{V}$ & -1 & 1 & 1 & -1 \\
    $F_{A}$ & 1 & 1 & -1 & -1 \\
    \hline
  \end{tabular},
\end{center}
where we have used the canonical anti-commutation relations $\{\lambda_{\pm,a},{\bar\lambda}_{\pm,b}\}=\delta_{ab}$. It looks like $F_{A}$ is not the symmetry of the Lagrangian (\ref{LGRM}). However, there is nothing wrong in considering it as an operator acting on the Hilbert of states of quantum mechanics, which is equivalent to rescaling the twisted superpotential $\widetilde{W}\rightarrow \zeta\widetilde{W}$. See \cite{Hori:2003ic}[Chapter 10.3.4] for more details. In our situation, this quantum mechanics comes from a two-dimensional quantum field theory. So when we put the two-dimensional theory on a finite volume, it indeed has a left unbroken finite $F_{A}$ symmetry and the center symmetry $\mathbb{Z}_{N}$ of the flavor group $SU(N)$. These extra symmetries are crucial to determine ground states in the interaction theory. The exact wave functions of ground states are, of course, difficult to compute. However, for our aims in this paper, we first consider the ground states in the limit where $\frac{\zeta}{\left|\Lambda\right|}$ goes to infinity. Thus the bosonic part of the wave functions is a delta function:
\begin{equation}\label{}
\prod^{k}_{a=1}\delta(x'_{a})\delta(y'_{a}).
\end{equation}
 A normalized wave function must have the ``fermion number" $k$ from the vector R-symmetry, so each vacuum should have $k$ ladder operators acting on it:
 \begin{equation}\label{}
  \prod^{l}_{b=1}{\bar\lambda}_{+,b}\prod^{m}_{c=1}{\bar\lambda}_{-,c}|0\rangle,
 \end{equation}
where the number of the ladder operators ${\bar\lambda}_{+}$, $l$, plus the number of ${\bar\lambda}_{-}$, $m$, should be $k$. However, since the axial $F_{A}$ is also a symmetry on a finite volume, when the far-infrared ground states map to the ones in the interacting quantum field theory, an extra constraint should be imposed
\begin{equation}\label{}
  l-m={\rm constant}.
\end{equation}
Since our study can be applied to any Grassmannian $Gr(k;N)$, there are two choices of the partition that are unique for any $k$: $l=k$, $m=0$, or $l=0$, $m=k$. In this paper, we choose $l=k$, $m=0$, and this means the ground state wave functions are
 \begin{equation*}
  \prod^{k}_{a=1}\delta(x'_{a})\delta(y'_{a}) \prod^{k}_{a=1}{\bar\lambda}_{+,a}|0\rangle.
\end{equation*}
Another evidence of these two choices can be seen from the two-dimensional point of view. In a topological-twisted LG model, some fermions are worldsheet Grassmann-odd scalars
\begin{equation}\label{}\nonumber
  {\bar \lambda}_{+;a}\leftrightarrow d\sigma_{a}.
\end{equation}
So the fermion number $k$ in all ground states can be expressed by a $k$-form:
\begin{equation}\label{}
  \prod^{k}_{a=1}{\bar\lambda}_{+,a}|0\rangle \leftrightarrow d\sigma_{1}\wedge\cdots\wedge d\sigma_{k}.
\end{equation}
The ground states locus are solutions of:
\begin{equation*}
  \left(\sigma_{a}\right)^{N}=(-)^{k-1}\left(\Lambda_{k}\right)^{N}.
\end{equation*}
 After choosing the bare theta angle for each $U(k)$ gauge theory to be $\theta_{k}=\theta+i\pi\left(k-1\right)$, the critical locus for any $k$ is actually can be understood by picking $k$ different sites from the $N$ homogenous sites on a circle defined by the dimensionless field space $\widehat{\sigma}=\sigma\Lambda^{-1}$, where $\Lambda=\mu e^{-r+i\theta}$,
\begin{equation}\label{}
  \widehat{\sigma}^{N}=1.
\end{equation}
So the sites are labeled by $\widehat{\sigma}_{j}=e^{\frac{2i\pi j}{N}}$. Notice the similarity between the delta function and the fermionic statistics, we propose the following enlarged fermionic operators ${\bar \lambda}_{+,j}$, and ${\bar \lambda}_{-,j}$, where $j\in\{1,\cdots,N\}$, to label the vacua. Up to an overall normalization, they are related to the original fermionic variables as
\begin{equation}\label{FDM}
  {\bar \lambda}_{\pm,j}=:\int d\widehat{{\bar\sigma}}d\widehat{\sigma} \delta^{2}\left(\widehat{\sigma}-\widehat{\sigma}_{j}\right){\bar \lambda}_{\pm}, \quad { \lambda}_{\pm,j}=: \int d\widehat{{\bar\sigma}}d\widehat{\sigma} \delta^{2}\left(\widehat{\sigma}-\widehat{\sigma}_{j}\right)\lambda_{\pm}
\end{equation}
and their canonical anti-commutation relations are
\begin{equation}\label{}
  \{{\bar\lambda}_{\pm,i}, \lambda_{\pm,j}\}=\delta_{ij}.
\end{equation}
They are non-local operators in field theory. The index $j$ labels the sites as well and is charged under the center group $\mathbb{Z}_{N}$ of the flavor symmetry $SU(N)$ with the charge $j$. The vector/axial R-symmetry can be defined similarly in terms of the new variables associated with $N$ sites, and they are
\begin{equation}\label{}
 F_{V}=-\sum^{N}_{j=1}\left({\bar\lambda}_{-,j}\lambda_{-,j}+{\bar\lambda}_{+,j}\lambda_{+,j}\right),\quad\quad F_{A}=\sum^{N}_{j=1}\left({\bar\lambda}_{+,j}\lambda_{+,j}-{\bar\lambda}_{-,j}\lambda_{-,j}\right).
\end{equation}

Therefore, the vacua can be described by a new set of variables as
\begin{equation}\label{NVISC}
  {\bar\lambda}_{-,1}\cdots{\bar\lambda}_{+,i_{1}}\cdots{\bar\lambda}_{+,i_{k}}\cdots{\bar\lambda}_{-,N}|0\rangle_{k}
\end{equation}
associated with a change in how operators act on a ``vacuum" in an isomorphic fashion: In old variables of gauge theory,
 \begin{equation}\label{}
   \lambda_{\pm,a}\mid 0\rangle=0, \quad {\rm for} \quad a\in \{1,\ldots,k\}.
 \end{equation}
 While in the new set of variables, we have
 \begin{equation}\label{}
   \lambda_{\pm,i}\mid 0\rangle_{k}=0, \quad {\rm for} \quad i\in \{1,\ldots,N\}.
 \end{equation}
 The number of all possible configuration in Eq.(\ref{NVISC}) is indeed the Witten index of $Gr(k;N)$:
  \begin{equation}\label{}\nonumber
    {N\choose k}.
  \end{equation}
  Superficially, ground states in Eq.(\ref{NVISC}) might have the ``fermion number" $N$, i.e. the vector R-symmetry. Nevertheless, the original ``fermion number" is $k$. In order to compensate for this difference, we assign a fermion number $k-N$ to the state $|0\rangle_{k}$ when we map it to the field theory. It is easy to see that the statistics of $|0\rangle_{k}$ is associated with the fermion number $k-N$. Although the physics is different, our manipulation looks similar to P.A.M. Dirac's operation to the negative energy states by introducing the so-called ``Dirac sea". In this convention, we can use the following notation to represent ground states, for example:
\begin{equation}\label{}
  |j_{1}j_{2}\rangle:=|-,\cdots,+_{j_{1}},\cdots,+_{j_{2}},\cdots,-\rangle_{2}.
\end{equation}
This notation is certainly not crucial for gauge theory itself. However, we will see later that it helps represent the connection between gauge theories and integrable systems, which in our paper will be formalized as a spin chain.

So far, we have defined the Hilbert space in the large volume limit. Recall that in higher dimensional quantum field theory, the domain wall fluctuation is significantly weaker than the perturbative quantum fluctuation at a sufficiently large distance. However, in a two-dimensional mass-gap theory, the perturbative quantum correction can be, in fact, much smaller than the domain wall amplitude at the intermediate scale, which allows us to truncate the quantum theory to a theory with only BPS-spectrum. We think this is the crucial reason why two dimensions intricately connect to an integrable system. To see this, we first estimate the two-point correlation function, up to an overall finite constant, in a (1+d)-dimensional mass-gap field theory. Take the physical mass as $m$, and then consider the limit where $m{\parallel} x{-}y{\parallel}\gg 1$:
\begin{equation}\label{}
  \langle\varphi(x)\varphi(y)\rangle\simeq\frac{m^{d-1}e^{-m{\parallel} x{-}y{\parallel}}}{\left(m{\parallel} x{-}y{\parallel}\right)^{\frac{d}{2}}}.
\end{equation}
The domain wall amplitude can be also expected, up to an overall finite constant, as
\begin{equation}\label{}
  \langle{\rm vacuum_{2}}|{\rm vacuum_{1}}\rangle\simeq e^{-HR}= e^{-TR^{d}},
\end{equation}
where $T$ is the tension of a domain wall, and it is RG-invariant for a BPS one. \\

$\bullet$ When $d>1$, if we identify $R$=${\parallel} x{-}y{\parallel}$, we find that the domain wall fluctuation, compared to the two-point function, is highly suppressed in the large volume limit. As said before, this is one physical reason for spontaneous symmetry breaking in higher-dimensional quantum field theory.

$\bullet$ When $d=1$, the physical mass $m$ of the spectrum in the perturbative region of the theory, which is called the perturbative spectrum in our paper, is much larger than the domain wall tension in a large volume:
\begin{equation}\label{}
  m\gg T.
\end{equation}
So at this intermediate scale, we have
\begin{equation}\label{}
  e^{-HR}\simeq e^{-TR}\gg\frac{e^{-m{\parallel} x{-}y{\parallel}}}{\left(m{\parallel} x{-}y{\parallel}\right)^{\frac{1}{2}}},
\end{equation}
The mass of the mode with a nonzero spatial momentum $p$ is heavier since the formula $m_{p}=\sqrt{m^{2}+p^{2}}$. So it is even more suppressed. This means that, at this scale, we only need to consider the dynamics of domain walls.

Therefore, one could discard the perturbative spectrum in a proper limit in a two-dimensional field theory. The ``LSZ"-like formula is, in fact, extremely elegant in this situation: \textit{The ground state wave functions defined in the far infrared can be mapped into an interacting theory by modifying them into a new basis that does not break the group $\mathbb{Z}_{N}$.} This physical constraint is due to the dynamics of domain walls since they are charged under $\mathbb{Z}_{N}$. So integrability could arise. For example, consider the $k=1$ case. The ground state wave functions can be represented as:
\begin{equation}\label{}
  A_{1}\sum^{N}_{j=1}\widehat{\sigma}^{j}|j\rangle_{1}.
\end{equation}
These wave functions are neutral under the finite group. For a general $k$, besides the finite group $\mathbb{Z}_{N}$, we should also impose a constraint due to the Weyl symmetry $S_{k}$. Thus, the ground state wave functions can be naturally constructed as
\begin{equation}\label{GRKGSWF}
  A_{k}\sum_{1\leq j_{1}<\cdots< j_{k}\leq N}\prod^{k}_{a=1}\sum_{{\cal W}\in S_{k}}\left(\widehat{\sigma}_{{\cal W}(a)}\right)^{j_{a}}|j_{1},\cdots ,j_{k}\rangle_{k}.
\end{equation}
where $ A_{k}$ is an overall normalization factor, $\widehat{\sigma}_{a}=\sigma_{a}\Lambda^{-1}$, ${\cal W}$ is an element of the Weyl group, and the periodic condition is
\begin{equation}\label{}
|j_{1},\cdots ,j_{k}\rangle_{k}=|j_{2},\cdots ,j_{k},j_{1}+N\rangle_{k}.
\end{equation}
The above periodic condition is associated with the equations $\left(\widehat{\sigma}_{a}\right)^{N}=1$ for each index $a$, which are the vacuum equations of each Grassmannian
\begin{equation}\label{}
 \left(\sigma_{a}\right)^{N}=(-1)^{k-1}\Lambda_{k}^{N}.
\end{equation}
Before going to the next, we want to point out that the energy barrier between two vacua will be finite at an intermediate scale. So the vacuum wave function changes from the delta function to a smearing wave packet on $\sigma$-space. It is certainly consistent with the fact that the domain walls are dynamic at a finite scale. \\

\noindent{\underline{Twisted Masses}}\\

So far, we have described the Hilbert space for Grassmannians without the twisted masses. When we turn on twisted masses with finite values, the center group, instead of $\mathbb{Z}_{N}$, is a generic $N$-order group. The vacuum equations
\begin{equation}\label{}
  \prod^{N}_{i=1}\left(\sigma_{a}+m_{i}\right)=\left(-1\right)^{k-1}\Lambda_{k}^{N}
\end{equation}
 still have $N$ solutions for each variable $\sigma_{a}$. Thus, there are still $N$ sites on the dimensionless field space $\widehat{\sigma}$. However, the length of the two nearest-neighbor sites is not translating invariant anymore in this situation. The procedure to define its Hilbert space is similar to before, and it is the state space of the so-called inhomogeneous spin chain.\\

\noindent{\underline{$T^{\ast}$Grassmannian}}\\

A similar construction applies to the superconformal field theory as well, if it has a sufficient global symmetry in the nonabelian gauge theory. In this situation, one can reduce the gauge theory to the Coulomb branch without missing any ground state.

To get the Hilbert space, we first turn on the generic twisted masses $m_{i}$ for the flavor symmetry $SU(N)$, where the conformal symmetry is broken, and the effective theory is essentially massive. Then the above analysis for the Grassmannian applies, which gives an inhomogeneous spin chain. If turning off $m_{i}$, domain walls are massless, where the conformal symmetry is restored, and the spin chain becomes a homogenous one.  Come back to the vacuum equations
\begin{equation}\label{}
  \left(\frac{\sigma_{a}-2su}{\sigma_{a}+2su}\right)^{N}=(-1)^{N}e^{-t}\prod_{a\neq b}\frac{\sigma_{a}-\sigma_{b}-2u}{\sigma_{a}-\sigma_{b}+2u}.
\end{equation}
We define the dimensionless variables as $\hat{\sigma}_{a}=\frac{\sigma_{a}}{2iu}$, while $e^{-t}$ is a dimensionless quantity that could not be simply absorbed by a field redefinition. Then the vacuum equations can be rewritten as
\begin{equation}\label{}
  \left(\frac{\hat{\sigma}_{a}+is}{\hat{\sigma}_{a}-is}\right)^{N}=(-1)^{N}e^{-t}\prod_{a\neq b}\frac{\hat{\sigma}_{a}-\hat{\sigma}_{b}+i}{\hat{\sigma}_{a}-\hat{\sigma}_{b}-i}.
\end{equation}
 Furthermore, they give
\begin{equation}\label{}
  \prod^{k}_{a=1}\left(\frac{\hat{\sigma}_{a}+is}{\hat{\sigma}_{a}-is}\right)^{N}=(-1)^{kN}e^{-kt}.
\end{equation}

As discussed in section \ref{Sec:2.1.1} for the $k=1$ case, the vacuum expectation value of
\begin{equation}\label{}
  \left(\frac{\hat{\sigma}_{a}+is}{\hat{\sigma}_{a}-is}\right)
\end{equation}
is charged under the center group $\mathbb{Z}_{N}$ of the flavor symmetry $SU(N)$. When we use the gauge-invariant operator
\begin{equation}\label{}
  e_{k}\left(\frac{\hat{\sigma}_{a}+is}{\hat{\sigma}_{a}-is}\right),
\end{equation}
it can label $N$ different sites. As in the Grassmannian case, there are also fermions ${\bar\lambda}_{\pm}$ at each site on the emergent spin chain. However, since there is an extra anomalous center symmetry $\mathbb{Z}_{2s}$ in the geometrical vacuum, it is natural to expect that there is one more index in fermions ${\bar\lambda}^{M}$ to reflect the fact that they are charged under $\mathbb{Z}_{2s}$ as well. Because of this index, one can put the spin operator ${\bar\lambda}^{M}{\bar\lambda}^{N}$ on the same site once the index $M$ is different from the index $N$. However, since the vacuum does not spontaneously break the global symmetry, one should regroup the composite operators such that they are neutral under $\mathbb{Z}_{2s}$. Thus, it has more Witten index for a larger $s$. Let us study each example more carefully.\\

$\bullet$ If $s=\frac{1}{2}$, the center group $\mathbb{Z}_{2s}$ is trivial. The ground state wave functions of this case can be proposed similarly. However, compared to the Grassmannian, there is an extra factor on the right hand side of the vacuum equations when $k>1$. The question is whether we can read off this extra factor from the wave functions. To start with, we first consider $k=1$, then the Hilbert space is
\begin{equation}\label{}
  A_{1}\sum^{N}_{j=1}\left(\frac{\sigma+\frac{i}{2}}{\sigma-\frac{i}{2}}\right)^{j}|j\rangle_{1}.
\end{equation}
To shorten the notation, we use
\begin{equation}\label{}
  e^{ip_{a}}:=\left(\frac{\sigma_{a}+\frac{i}{2}}{\sigma_{a}-\frac{i}{2}}\right),
\end{equation}
and the physical meaning of this notation will be discussed in the following section. Next, we consider the $U(2)$ case. The proposed Hilbert space is then
\begin{equation}\label{TGRHS}
  A_{2}\sum_{1\leq j_{1}< j_{2}\leq N}\left(e^{ip_{1}j_{1}}e^{ip_{2}j_{2}}+S_{21}e^{ip_{2}j_{1}}e^{ip_{1}j_{2}}\right)|j_{1}<j_{2}\rangle_{2},
\end{equation}
where the scattering factor is
\begin{equation}\label{}
  S_{12}=\frac{\hat{\sigma}_{1}-\hat{\sigma}_{2}+i}{\hat{\sigma}_{1}-\hat{\sigma}_{2}-i},
\end{equation}
and the twisted boundary condition of $N$ sites
\begin{equation}\label{OSBC}
  |j_{2}<j_{1}+N\rangle_{2}:=(-1)^{N}q^{-1}|j_{1}<j_{2}\rangle_{2}.
\end{equation}
Now we check that Eq.(\ref{TGRHS}) is invariant under the Weyl symmetry $S_{2}$, which acts on the field variables as $\sigma_{1}\leftrightarrow\sigma_{2}$. Eq.(\ref{TGRHS}) becomes
\begin{equation}\label{}
   A_{2}\sum_{1\leq j_{1}< j_{2}\leq N}\left(e^{ip_{2}j_{1}}e^{ip_{1}j_{2}}+S_{12}e^{ip_{1}j_{1}}e^{ip_{2}j_{2}}\right)|j_{1}<j_{2}\rangle_{2}.
\end{equation}
By using the fact $S_{12}S_{21}=1$, then we have
\begin{equation}\label{}
   A'_{2}\sum_{1\leq j_{1}< j_{2}\leq N}\left(S_{21}e^{ip_{2}j_{1}}e^{ip_{1}j_{2}}+e^{ip_{1}j_{1}}e^{ip_{2}j_{2}}\right)|j_{1}<j_{2}\rangle_{2},
\end{equation}
where $A'_{2}=A_{2}S_{12}$. So it is invariant under the Weyl symmetry up to an overall factor. From the boundary condition (\ref{OSBC}), one can derive the vacuum equations
\begin{equation}\label{}
  e^{ip_{1}N}=(-1)^{N}q S_{12},\quad\quad e^{ip_{2}N}=(-1)^{N}q S_{21}.
\end{equation}
If $(-1)^{N}q=\pm1$, one can show that $p_{a}$ is real and $S_{12}$ is a pure phase that can be written as $e^{i\theta(p_{1},p_{2})}$, where $\theta(p_{1},p_{2})$ is a real function. However, if $(-1)^{N}q\neq \pm1$, $p_{a}$ is a complex number, and $S_{12}$ is not a pure phase. We call it the \textit{open spin chain} in section \ref{Sec:3}. For a general $k$, we have the following proposed Hilbert space
\begin{equation}\label{GTGRH}
    A_{k}\sum_{1\leq j_{1}<\cdots <j_{k}\leq N}\sum_{{\cal W}\in S_{k}}S_{{\cal W} \left(a\right)}e^{\sum^{k}_{a=1}ip_{{\cal W} \left(a\right)}j_{a}}|j_{1}<\cdots<j_{k}\rangle_{k},
\end{equation}
 where $S_{{\cal W}(a)}=\prod_{ba}S_{ba}$, and each factor $S_{ba}$ is due to a permutation (elastic scattering) between the momentum $p_{a}$ and $p_{b}$. The twisted boundary condition
 \begin{equation}\label{}
   |j_{2}<\cdots<j_{1}+N\rangle_{k}:=(-)^{N}q^{-1}|j_{1}<\cdots j_{k}\rangle_{k}.
 \end{equation}
 Up to the ambiguity of an overall normalization factor, one can show that Eq.(\ref{GTGRH}) is invariant under the Weyl group. Furthermore, from the twisted boundary condition, one can derive the Coulomb branch vacuum equations of GLSM for $T^{\ast}Gr(k;N)$, and they are
 \begin{equation}\label{}
   e^{ip_{a}N}\prod_{b\neq a}S_{ba}=(-1)^{N}q.
 \end{equation}
Finally, we want to comment that the presentation of wave function in our paper is not the only option. One can, of course, choose a different basis such that the wave function can be factorized into a product, which would manifest the so-called \textit{Algebraic Bethe Ansatz} in this manner.

 $\bullet$ If $s=1$, the center group $\mathbb{Z}_{2}$ is nontrivial. The fermion on each site, ${\bar\lambda}^{M}_{+}$, have an extra index $M=\{0, 1\}$. The center group $\mathbb{Z}_{2}$ acts on the fermion ${\bar\lambda}^{M}_{+}$ by ${\bar\lambda}^{0}_{+}\mapsto {\bar\lambda}^{1}_{+}$. One should focus on the $\mathbb{Z}_{2}$ neutral operators, since the vacuum does not break this $\mathbb{Z}_{2}$ symmetry. The possible neutral operators for this case at each site are 1, ${\bar\lambda}^{0}_{+}+{\bar\lambda}^{1}_{+}$ and ${\bar\lambda}^{0}_{+}{\bar\lambda}^{1}_{+}+{\bar\lambda}^{1}_{+}{\bar\lambda}^{0}_{+}$. So the number of all possible states is $3^{N}$.

 Now let us focus on the specific example $T^{\ast}Gr(k;N)$. If $k=1$, the ground states are
 \begin{equation}\label{}
   |j^{\langle1\rangle}\rangle_{1},
 \end{equation}
 where the index $\langle1\rangle$ means the vacuum is created by the operator ${\bar\lambda}^{0}_{j, +}+{\bar\lambda}^{1}_{j, +}$. So we have $N$ ground states for $k=1$.  If $k=2$, we have two types of the ground states
 \begin{equation}\label{}
   |j_{1}^{\langle1\rangle}< j_{2}^{\langle1\rangle}\rangle_{2}, \quad {\rm and}\quad |j^{\langle2\rangle}\rangle_{2},
 \end{equation}
 where the notation $j^{\langle2\rangle}$ means the vacuum is constructed from the operator ${\bar\lambda}^{0}_{l, +}{\bar\lambda}^{1}_{l,+}+{\bar\lambda}^{1}_{l,+}{\bar\lambda}^{0}_{l,+}$.  This says we have
 \begin{equation}\label{}
   {N \choose 2}+{N \choose 1}
 \end{equation}
 ground states. The Hilbert space can be proposed as
 \begin{equation}\label{}
     A_{2}\left(\sum_{1\leq j^{\langle1\rangle}_{1}< j^{\langle1\rangle}_{2}\leq N}\left(e^{ip_{2}j^{\langle1\rangle}_{1}}e^{ip_{1}j^{\langle1\rangle}_{2}}+S_{12}e^{ip_{1}j^{\langle1\rangle}_{1}}e^{ip_{2}j^{\langle1\rangle}_{2}}\right)|j^{\langle1\rangle}_{1}<j^{\langle1\rangle}_{2}\rangle_{2}+C\sum_{1\leq j^{^{\langle2\rangle}}\leq N}e^{i(p_{1}+p_{2})j^{\langle2\rangle}}|j^{^{\langle2\rangle}}\rangle_{1}\right).
 \end{equation}
 The notation $C$ stands for a contact term, so it is scheme dependent and can be changed if we modify the theory in the UV. The Weyl symmetry, in fact, changes the coefficients $A_{2}$ and $C$ by
 \begin{equation}\label{}
   A_{2}\mapsto A'_{2}=A_{2}S_{21},\quad\quad C\mapsto C'=CS_{12}.
 \end{equation}
So the Weyl symmetry is associated with a possible ambiguity of the UV contact term. The twisted boundary conditions are
\begin{equation}\label{}
  |j^{\langle1\rangle}_{2}<j^{\langle1\rangle}_{1}+N\rangle_{2}=(-1)^{N}q^{-1}|j^{\langle1\rangle}_{1}<j^{\langle2\rangle}_{2}\rangle_{2},\quad\quad |j^{^{\langle2\rangle}}+N\rangle_{1}=(-1)^{2N}q^{-2}|j^{^{\langle2\rangle}}\rangle_{1}.
\end{equation}
 From these boundary conditions, one can derive the vacuum equations
 \begin{equation}\label{}
   e^{ip_{1}N}=(-1)^{N}qS_{12},\quad\quad  e^{ip_{2}N}=(-1)^{N}qS_{21},\quad\quad  e^{i(p_{1}+p_{2})N}=q^{2},
 \end{equation}
 where the last equation is not independent and can be derived from the first two.

 For a general $k\leq N$, the analysis is similar. It has the number of ground states
 \begin{equation}\label{}
   \sum^{[\frac{k}{2}]}_{l=0}{N \choose l}{N-l \choose k-2l},
 \end{equation}
 where $[\frac{k}{2}]$ is the integer part of $\frac{k}{2}$. Our framework predicts that one can consider the gauge theory with $2N\geq k>N$ as well since the adjoint matter field is sufficient to Higgsing the gauge field. The upper bound $2N$ can be seen from the fact that the maximal number of ${\bar\lambda}_{+}$ is $2n$ through the operator $\prod^{N}_{l=1}\left({\bar\lambda}^{0}_{l, +}{\bar\lambda}^{1}_{l,+}+{\bar\lambda}^{1}_{l,+}{\bar\lambda}^{0}_{l,+}\right)$. The number of ground states for the case $2N\geq k>N$ is
 \begin{equation}\label{}
   \sum^{[\frac{k}{2}]}_{l'\geq k-N }{N \choose l'}{N-l' \choose k-2l'}.
 \end{equation}
 One can show that the number of ground states for $k$ is the same as the number of ground states for the $2N-k$ case. Thus, we conjecture that there could be a Seiberg-dual between the $U(k)$ gauge theory and the associated $U(2N-k)$ one with proper matter content. The detailed study of these gauge theories is left to future work.

 $\bullet$ If s$>$1, the discussion is more complicated but very similar to the $s=1$ case. We leave the details to the interested reader. \\

 \noindent{\underline{Quiver Gauge Theories}}\\

A complete study of GLSMs for quiver gauge theories does not appear in the literature yet. However, we expect that the scattering factor in the vacuum equations on the Coulomb branch should be
 \begin{equation}\label{QGLSM}
   S^{n_{1},n_{2}}=\frac{\sigma^{n_{1}}_{a}-\sigma^{n_{2}}_{b}-\frac{i}{2}\left(DA\right)_{n_{1}n_{2}}}{\sigma^{n_{1}}_{a}-\sigma^{n_{2}}_{b}+\frac{i}{2}\left(DA\right)_{n_{1}n_{2}}},
 \end{equation}
where $DA$ denotes the symmetric Cartan matrix encoded into the Dynkin diagram of the corresponding algebra. For example, $SU(L+1)\simeq {\rm A_{L}}$ group has the Cartan matrix
\begin{equation*}
  DA=\left(
  \begin{array}{cccc}
    +2 & -1 &  &  \\
    -1 & \ddots & \ddots & \\
     & \ddots & \ddots & -1\\
     & & -1 & +2 \\
  \end{array}
\right).
\end{equation*}\\

 \noindent{\underline{Comment On Other Cases}}\\

 So far, we have discussed the cases all have the structure that
 \begin{equation}\label{EKSC}
   \left(e_{k}\left(\sigma\right)\right)^{N}={\rm Constant},
 \end{equation}
 where $e_{i}(\sigma)$ is the $i$-th elementary symmetric polynomial. One can also show that the GLSMs for Flag manifolds such as $F(k_{1},k_{2}; N)$ \cite{Donagi:2007hi} share the same structure as the above equation. However, some other target spaces do not pose this structure. For example, see the vacuum equations of GLSM for a Lagrangian Grassmannian $LG(N,2N)$ \cite{Gu:2020oeb}
 \begin{equation}\label{VELG}
   \left(\sigma_{a}\right)^{2N}=q\prod_{b\neq a}\left(\sigma_{a}+\sigma_{b}\right), \quad\quad \left(\sigma_{a}\right)^{2}\neq \left(\sigma_{b}\right)^{2}.
 \end{equation}
 One can easily observe that the above equations do not have the structure of Eq.(\ref{EKSC}). In order to incorporate the structure (\ref{EKSC}) in the Lagrangian Grassmannian case, we first symmetrize Eq.(\ref{VELG}). For example, when $N=2k+1$, it can be expanded as
 \begin{equation}\label{LGVE1}
   \left(\sigma_{a}\right)^{4k+2}=q\left(\left(\sigma_{a}\right)^{2k}+\left(\sigma_{a}\right)^{2k-2}e_{2}(\sigma)+\cdots+e_{2k-2}(\sigma)\left(\sigma_{a}\right)^{2}+e_{2k}(\sigma)\right).
 \end{equation}
 Similarly, for $N=2k$, it can be rewritten as
 \begin{equation}\label{LGVE2}
\left(\sigma_{a}\right)^{4k}=q\left(\left(\sigma_{a}\right)^{2k-2}e_{1}(\sigma)+\left(\sigma_{a}\right)^{2k-4}e_{3}(\sigma)+\cdots+e_{2k-3}(\sigma)\left(\sigma_{a}\right)^{2}+e_{2k-1}(\sigma)\right).
 \end{equation}
  From the above two polynomials, it is easy to find that Eq.(\ref{LGVE1}) and Eq.(\ref{LGVE1}) can be mapped to the ones of the equivariant Grassmannian with specific parameters.  To see this, we turn on the twisted masses for the Grassmannian:
  \begin{equation}\label{}
    \prod^{2N}_{i=1}\left(\sigma_{a}+m_{i}\right)=(-1)^{N-1}\Lambda.
  \end{equation}
  Rewriting the above as
  \begin{equation}\label{EGRVE}
   \left(\sigma_{a}\right)^{2N}+\left(\sigma_{a}\right)^{2N-1}e_{1}\left(m_{i}\right)+\cdots+\left(\sigma_{a}\right)e_{2N-1}\left(m_{i}\right)+e_{2N}\left(m_{i}\right)=(-1)^{N-1}\Lambda.
  \end{equation}
  So if we identify the coefficients of (\ref{EGRVE}) with the ones in Eq. (\ref{LGVE1}) or (\ref{LGVE2}) and further restrict to the locus where $\left(\sigma_{a}\right)^{2}\neq \left(\sigma_{b}\right)^{2}$, then we can reproduce the vacuum equations of a Lagrangian Grassmannian from the vacuum equations of the equivariant Grassmannian. Based on this, we conjecture that the spin chain for the Lagrangian Grassmannian corresponds to an inhomogeneous spin chain for the Grassmannian. A detailed investigation would be found in \cite{Gu:20xx}.

\subsection{3d ${\cal N}$=2 Chern-Simons Matter Theory}\label{Sec:2.2}
In the three-dimensional gauge theory, we have an extra ingredient, which is the so-called Chern-Simons interaction:
\begin{equation}\label{}
 {\cal S}^{{\cal N}=2}_{{\rm CS}}=\frac{k}{4\pi} \int d^{3}x {\rm Tr} \left(AdA+\frac{2}{3}A^{3}-{\bar \lambda}\lambda+2D\sigma\right).
\end{equation}
The Chern-Simons interaction affects low-energy physics with a new phase: the topological gauge theory \cite{Witten:1999ds}. For our purposes in this paper, we mainly study the 3d ${\cal N}$=2 Chern-Simons matter theory on space-time $\mathbb{R}^{2}\times S^{1}$. Our motivation relies on the fact that the connection to the integrable system is essentially the two-dimensional quantum field theory as we discussed in the previous section.\\

\subsubsection{Grassmannain}
We first discuss the 3d Chern-Simons theory for Grassmannian. There is no theta angle in the odd-dimensional gauge theory, so the phases of 3d CS theory are parameterized by the one real dimensional FI parameter $\xi$. However, if put the theory on space-time $\mathbb{R}^{2}\times S^{1}$, a theta angle would emerge. The Chern-Simons interaction can affect low-energy physics. \\

\noindent{\underline{Quantum K-Theory With Level Structure}}\\

Let us first choose the bare Chern-Simons level $k=-\frac{N}{2}$, then if $\xi\gg0$, the low energy effective theory is a three-dimensional nonlinear sigma model on Grassmannian;  if $\xi\ll0$, the low energy effective theory, after integrating out the massive fermions, is the Chern-Simons theory $U(k)_{N-k}$, where the subscript $N-k$ is the level. If we decompose the Lie algebra $U(k)=SU(k)\times U(1)$, the level for $SU(k)$ sector is $N-k$ and for the $U(1)$ sector is $N$. If further put the theory on the space-time $\mathbb{R}^{2}\times S^{1}$, where the radius of $S^{1}$ is $R$, it reduces to a two-dimensional gauge theory with an infinite number of massive fields from the KK-reduction. The parameter becomes a complexified one: $t=r-i\theta$, where $\theta$ is an emergent theta angle from the KK-reduction. The 3d parameter relates to the 2d FI-parameter by $q_{3d}=\left(-2\pi R\right)^{N}\Lambda^{N}$, where $q_{3d}=e^{-\xi}$, and $\Lambda$ is the dynamical scale of the two-dimensional gauge theory.

If $r\gg0$, the low energy effective theory is a K-theoretic lifting of the two-dimensional NLSM on a Grassmannian \cite{Kapustin:2013hpk}; if $r\ll0$, the effective theory is a K-theoretic version of the twisted effective superpotential:
\begin{eqnarray}
   R\star\widetilde{W}^{3d}_{{\rm eff}} &=& \frac{1}{2}k_{SU(k)}\sum^{k}_{a=1}\left(\ln X_{a}\right)^{2}+\frac{k_{U(1)}-k_{SU(k)}}{2k}\left(\sum^{k}_{a=1}\ln X_{a}\right)^{2} \\\nonumber
   &&+\ln\left((-1)^{k-1}q_{3d}\right)\sum^{k}_{a=1}\ln X_{a}+N\sum^{k}_{a=1}{\rm Li_{2}}\left(X_{a}\right)+\frac{N}{4}\sum^{k}_{a=1}\left(\ln X_{a}\right)^{2},\\\nonumber
  &=& \left(\ln(-1)^{k-1}q_{3d}\right)\sum^{k}_{a=1}\ln X_{a}+N\sum^{k}_{a=1}{\rm Li_{2}}\left(X_{a}\right),
\end{eqnarray}
where $X_{a}=e^{2\pi R\Sigma_{a}}$, and we have split the level $k_{U(k)}$ into two parts: $k_{U(1)}$ and $k_{SU(k)}$. The vacuum equations are
\begin{equation}\label{}
  \left(1-x_{a}\right)^{N}=(-1)^{k-1}q_{3d},\quad\quad {\rm where}\quad\quad x_{a}\neq x_{b},\quad {\rm if} \quad a\neq b.
\end{equation}
If we map the shifted Wilson loop operators $z_{a}=1-x_{a}$ to the Chern roots $\sigma_{a}$, the above vacuum equations are essentially the same as the vacuum equations for the quantum cohomology of Grassmannian. It is not surprising. As first observed in \cite{Witten:1993xi} by Witten that the quantum cohomology of Grassmannian connects to the Verlinde algebra of $U(k)$ via a smooth path on the parameter space. Then, its K-theoretic version of this correspondence has been discussed in \cite{Kapustin:2013hpk} by Kapustin and Willet. See \cite{Ruan2018} also for a mathematical definition of the quantum K-theory with a level structure by Ruan-Zhang. Moreover, the BPS spectrum and Hilbert space are similar to the two-dimensional situation, with the only modification being to replace the field variables $\sigma_{a}$ by $z_{a}$. The analysis of the equivariant situation is also resembling.\\

\noindent{\underline{Quantum K-Theory}}\\

If we choose a different gauge-invariant Chern-Simons level such as $k_{U(1)}=-\frac{N}{2}$ and $k_{SU(k)}=k-\frac{N}{2}$ \cite{Jockers:2019lwe,Ueda:2019qhg,Gu:2020zpg,Gu:2021yek,Gu:2022yvj}, then the low energy effective theory would be different. If $r\gg 0$, it is the NLSM on the Grassmannian, but its topological sector corresponds to the Givental-Lee's quantum K theory; if $r\ll 0$, we have the twisted effective superpotential
\begin{equation}\label{}
  R\star\widetilde{W}^{3d}_{{\rm eff}}=\frac{k}{2}\sum^{k}_{a=1}\left(\ln X_{a}\right)^{2}-\frac{1}{2}\left(\sum^{k}_{a=1}\ln X_{a}\right)^{2}+\left(\ln(-1)^{k-1}q_{3d}\right)\sum^{k}_{a=1}\ln X_{a}+N{\rm Li_{2}}(X_{a}).
\end{equation}
The vacuum equations are
\begin{equation}\label{KGRVE}
  \left(1-x_{a}\right)^{N}=(-)^{k-1}q_{3d}\frac{\left(x_{a}\right)^{k}}{\prod^{k}_{b=1}x_{b}}.
\end{equation}
The extra homogenous factor on the right-hand side indicates that the Hilbert space of the interacting theory is more complicated than the previous choice: $k_{U(1)}=k_{SU(k)}=-\frac{N}{2}$. The Hilbert space can be constructed similarly as in Eq.(\ref{GTGRH})
\begin{equation}\label{GTGRH2}
    A_{k}\sum_{1\leq j_{1}<\cdots <j_{k}\leq N}\sum_{{\cal W}\in S_{k}}S_{{\cal W} \left(a\right)}e^{\sum^{k}_{a=1}ip_{{\cal W} \left(a\right)}j_{a}}|j_{1}<\cdots<j_{k}\rangle_{k},
\end{equation}
where $S_{{\cal W} \left(a\right)}=\prod_{ba}S_{ba}$ and
\begin{equation}\label{}
  S_{ab}=-\frac{x_{a}}{x_{b}}.
\end{equation}
If we identify the variables as
\begin{equation}\label{QKCSM}
  e^{ip_{a}}:=z_{a}.
\end{equation}
However, the ``momentum" $p_{a}$ is a complex one. The boundary condition is
\begin{equation}\label{}
  |j_{2}<\cdots<j_{k}<j_{1}+N\rangle_{k}=\left(q_{3d}\right)^{-1}|j_{1}<\cdots<j_{k}\rangle_{k}.
\end{equation}

The (anti)-periodic boundary is satisfied if we set $q_{3d}=\pm1$. However, unlike $T^{\ast}Gr(k;N)$, the momentum $p_{a}$ in Eq.(\ref{QKCSM}) is always complex even though $q_{3d}=\pm1$ because the scattering factor $S_{ab}$ is not a pure phase factor in this case.

Finally, we want to mention that one can also have other gauge-invariant choices of Chern-Simons level besides the above two quantum K-theories. We propose that those quantum K theories correspond to in-homogenous XX spin chains.\\

\noindent{\underline{Exact Theory On Space-time $\mathbb{R}\times S^{1}$}}\\

Each KK-mode can be understood as a 2d nonabelian gauge theory with a twisted mass $\frac{in}{R}$. By using the exact result of the 2d nonabelian gauge theory,  one can obtain a Landau-Ginzberg theory on space-time $\mathbb{R}\times S^{1}$ by summing all KK-modes. This theory includes the field configuration of domain walls, although non-local. It would be then interesting to see what could happen if we take the decompactification limit
of the radius $R$.\\

\subsubsection{The $T^{\ast}Gr(k;N)$ Sigma Model}\label{KTGR}

  The condition of 3d Chern-Simons-matter theory to be renormalizable requires that the power of $X$ in the superpotential (\ref{TGRS}) satisfies $1\leq2s\leq2$. With a proper Chern-Simons level, one can do the following twisting,
  \begin{equation}\label{}
      \left(1-x^{-1}\right)\mapsto x\left(1-x^{-1}\right)=x-1,
  \end{equation}
  toward a negative charge-one bundle in K-theory. The K-theoretic vacuum equations for $T^{\ast}Gr(k;N)$ have been investigated in \cite{Nekrasov:2009uh}, and they are
\begin{equation}\label{CSVE}
  \left(\frac{\sin\left(\pi \widetilde{R}\left(\widehat{\sigma}_{a}+is\right)\right)}{\sin\left(\pi \widetilde{R}\left(\widehat{\sigma}_{a}-is\right)\right)}\right)^{N}=\widetilde{q}\prod^{k}_{b\neq a}\left(\frac{\sin\left(\pi \widetilde{R}\left(\widehat{\sigma}_{a}-\widehat{\sigma}_{b}+i\right)\right)}{\sin\left(\pi \widetilde{R}\left(\widehat{\sigma}_{a}-\widehat{\sigma}_{b}-i\right)\right)}\right),
\end{equation}
where $\widetilde{R}=2Ru$, $\widehat{\sigma}_{a}=\frac{\sigma_{a}}{2iu}$, and $\widetilde{q}=qe^{2\pi\widetilde{R}(Ns+1-k)}(-1)^{N}$. The Hilbert space can be constructed similarly by identifying
\begin{equation}\label{}
  e^{ip_{a}}=\frac{\sin\left(\pi \widetilde{R}\left(\widehat{\sigma}_{a}+is\right)\right)}{\sin\left(\pi \widetilde{R}\left(\widehat{\sigma}_{a}-is\right)\right)}.
\end{equation}
From Eq.(\ref{CSVE}), one can find that
\begin{equation}\label{}
  e^{i\left(\sum^{k}_{a=1}p_{a}\right)N}=\widetilde{q}^{k}.
\end{equation}
So the Hilbert space of this case can be constructed similarly to before with a new phase factor
\begin{equation}\label{}
  S_{ab}=\frac{\sin\left(\pi \widetilde{R}\left(\widehat{\sigma}_{a}-\widehat{\sigma}_{b}+i\right)\right)}{\sin\left(\pi \widetilde{R}\left(\widehat{\sigma}_{a}-\widehat{\sigma}_{b}-i\right)\right)}.
\end{equation}

\subsection{4d ${\cal N}=1$ Gauge Theory}\label{Sec:2.3}
Compared to the lower-dimensional quantum field theories, the gauge anomaly cancellation of the four-dimensional gauge theory imposes an extra constraint on the field content. Assigning the gauge charge $q_{i}$ to each matter field under each gauge group $U(1)$, then the gauge anomaly cancellation says $\sum_{i}\left(q_{i}\right)^{3}=0$.  If we consider a fundamental matter, the simplest choice to cancel the gauge anomaly is to pair it with an anti-fundamental one. Therefore, the Grassmannian target space is not allowed in the four-dimensional quantum field theory since it only has the fundamental matters.  While the $T^{\ast}Gr(k;N)$ sigma model, of course, can be defined in four dimensions. In this section, we mainly focus on the gauge theory with paired fundamental and anti-fundamental matters and a matter in the adjoint representation. A renormalizable superpotential is $W={\rm Tr}\widetilde{\Phi}X\Phi$. \\

\subsubsection{The $T^{\ast}Gr(k;N)$ Sigma Model}
Nekrasov and Shatashvili studied the 4d ${\cal N}=1$ gauge theory on space-time $\mathbb{R}^{2}\times \mathbb{T }^{2}$ in \cite{Nekrasov:2009uh}. We denote $\tau$ to be the complex structure of the torus $\mathbb{T }^{2}$. The vacuum equations are then
\begin{equation}\label{4VEGR}
  \left(\frac{\Theta_{1}\left(i\pi \widetilde{R}\left(\widehat{\sigma}_{a}+is\right)\right)}{\Theta_{1}\left(i\pi \widetilde{R}\left(\widehat{\sigma}_{a}-is\right)\right)}\right)^{N}=\widetilde{q}\prod_{b\neq a}\left(\frac{\Theta_{1}\left(i\pi \widetilde{R}\left(\widehat{\sigma}_{a}-\widehat{\sigma}_{b}+i\right)\right)}{\Theta_{1}\left(i\pi \widetilde{R}\left(\widehat{\sigma}_{a}-\widehat{\sigma}_{b}-i\right)\right)}\right),
\end{equation}
where:
\begin{equation}\label{}
  \Theta_{1}\left(\alpha\right)=-iq^{\frac{1}{8}}\left(e^{\alpha}-e^{-\alpha}\right)\prod^{\infty}_{m=1}\left(1-q^{m}\right)\left(1-q^{m}e^{2\alpha}\right)\left(1-q^{m}e^{-2\alpha}\right).
\end{equation}\\

If consider a 4d ${\cal N}=1$ $U(k)$ gauge theory for just $N$ fundamental and anti-fundamental matters with the same masses, the vacuum equations are simpler:
\begin{equation}\label{4VEGR2}
  \left(\frac{\Theta_{1}\left(i\pi \widetilde{R}\left(\widehat{\sigma}_{a}+is\right)\right)}{\Theta_{1}\left(i\pi \widetilde{R}\left(\widehat{\sigma}_{a}-is\right)\right)}\right)^{N}=\widetilde{q}.
\end{equation}

\noindent{\underline{Axial R-Symmetry}}\\

It is known that the axial R-symmetry is related to the isometry of the extra dimensions. For a generic torus $\mathbb{T}^{2}$, the isometry is $\mathbb{Z}_{2}\rtimes \mathbb{T}^{2}$. The $\mathbb{Z}_{2}$ is generated by the antipodal map \footnote{If we realize the flat torus $\mathbb{T}^{2}=\mathbb{R}^{2}/\Gamma$ for some plane lattice
 \begin{equation}\label{}\nonumber
   \Gamma_{e_{1},e_{2}}=\left\{z_{1}e_{1}+z_{2}e_{2}: z_{1}, z_{2}\in \mathbb{Z}\right\},
\end{equation}
 then one can show that the isometry of the torus is $\left(O(2)\cap {\rm Aut}(\Gamma)\right)\rtimes \mathbb{T}^{2}$, where ${\rm Aut}(\Gamma)$ is the automorphism group of the lattice $\Gamma$.}. However, the translation symmetry $\mathbb{T}^{2}$ is the reason for the ``twisted mass" $\widetilde{m}=\frac{i(1+\tau)}{R}$\cite{Nekrasov:2009uh}. And the Kaluza-Klein modes with momentum $m,n \in \mathbb{Z}$, would have the corresponding twisted mass
 \begin{equation}\label{}
   \widetilde{m}_{m,n}=\frac{i}{R}\left(m+n\tau\right).
\end{equation}
 Therefore, the gauge theory does not have a conserved $U(1)$ axial R-symmetry, but only the left $\mathbb{Z}_{2}$ one. While the vector R-symmetry $U(1)$ is conserved and comes from the $U(1)$ R-symmetry of the original 4d ${\cal N}$=1 gauge theory. It suggests that the domain walls of this model are different from the ones in the usual 2d GLSM. In this paper, however, we will use symmetry to guide our understanding of the connection between gauge theories and integrable systems for this case. While we leave the full investigation of the spectrum to a future study \cite{Gu:20xx}. Finally, we want to point out that there could be a special torus with a larger finite group ${\rm Aut}(\Gamma)$, and their corresponding integrable system could be slightly different from the $\mathbb{Z}_{2}$ case.

 We end this section by mentioning that it is possible to have a more generic Heisenberg spin chain from gauge theory if one could formalize it on space-time $\mathbb{R}^{2}\times C$, where $C$ is a non-orientable surface such as Klein bottle.

\subsection{Other paths to a ${\cal N}$=(2,2) gauge theory}\label{Sec:2.4}

In this section, we briefly review several other routes to an ${\cal N}$=(2,2) gauge theory. They are all related to the 4d ${\cal N}$=2 theory. None of these results are new.\\

\noindent{ {\underline{4d ${\cal N}=2$ Gauge Theory With The $\Omega$-background $\mathbb{R}^{2}\times \mathbb{R}^{2}_{\varepsilon}$}}}\\

Nekrasov and Shatashvili \cite{Nekrasov:2009rc} considered the four-dimensional ${\cal N}=2$ gauge theory on space-time $\mathbb{R}^{2}\times \mathbb{R}^{2}_{\varepsilon}$. If we denote the coordinates on $\mathbb{R}^{2}_{\varepsilon}$ as $(x^{2},x^{3})$, then the $U(1)$ rotation symmetry of this space is generated by the vector field
 \begin{equation}\label{}
  U=x^{2}\partial_{3}-x^{3}\partial_{2}.
\end{equation}
 The $\Omega$ deformation on $\mathbb{R}^{2}_{\varepsilon}$ is generated by $U$, and the associated bare Lagrangian of the 4d gauge theory is deformed in this background. Although the Poincare symmetry is broken in $\mathbb{R}^{2}_{\varepsilon}$, it still has a two-dimensional ${\cal N}$=(2,2) super-Poincare symmetry in the $\mathbb{R}^{2}$ directions.

 The strategy in \cite{Nekrasov:2009rc} to get the 2d twisted effective superpotential from a prepotential of the four-dimensional gauge theory is the following: First, consider the four-dimensional theory in a general $\Omega$-background $\mathbb{R}^{2}_{\widetilde{\varepsilon}}\times \mathbb{R}^{2}_{\varepsilon}$, with the first rotation parameter $\widetilde{\varepsilon}$ to be nonzero as well. Following \cite{Nekrasov:2002qd}, one can then derive an exact prepotential
 \begin{equation}\label{4DEP}
   S^{{\rm 4d}}_{{\rm eff}}=\int {\cal F}^{{\rm 4d} }\left(\sigma;\varepsilon,\widetilde{\varepsilon}\right)+\left\{Q, \ldots\right\}
 \end{equation}
where we denote collectively by $\sigma$ for all the vector multiplet scalars as well as background scalars, such as the masses of matter fields. Then the two-dimensional twisted effective superpotential can be obtained similarly as above in the $\Omega$-background
\begin{equation}\label{2DTE}
  S^{{\rm 2d}}_{{\rm eff}}=\int \widetilde{W}_{{\rm eff}}\left(\sigma; q ;\widetilde{\varepsilon}\right)+\left\{Q,\ldots\right\}.
\end{equation}
Following \cite{Nekrasov:2009rc}, modulo $Q$-exact terms, we have
\begin{equation}\label{EP420}
  \int_{\mathbb{R}^{4}} {\cal F}^{{\rm 4d} }\left(\sigma;\varepsilon,\widetilde{\varepsilon}\right)=\frac{1}{\varepsilon}\int_{\mathbb{R}^{2}} {\cal F}^{{\rm 2d} }\left(\sigma;\varepsilon,\widetilde{\varepsilon}\right)=\frac{1}{\widetilde{\varepsilon}\varepsilon}{\cal F} \left(\sigma;\varepsilon,\widetilde{\varepsilon}\right).
\end{equation}
The partition function of an effective theory is
\begin{equation}\label{}
  {\cal Z}\left(\sigma;q;\varepsilon,\widetilde{\varepsilon} \right)=\exp\left(\frac{1}{\varepsilon\widetilde{\varepsilon}}{\cal F} \left(\sigma;\varepsilon,\widetilde{\varepsilon}\right)\right).
\end{equation}
Finally, from the formulas (\ref{4DEP}), (\ref{2DTE}), and (\ref{EP420}), one can observe the connection between the 2d twisted effective superpotential and the 4d prepotential by
\begin{equation}\label{2df4d}
  \widetilde{W}_{{\rm eff}}(a;q;\varepsilon)=\lim_{\widetilde{\varepsilon}\rightarrow 0}\left[\widetilde{\varepsilon}\log{\cal Z}\left(\sigma;q;\varepsilon,\widetilde{\varepsilon} \right)\right].
\end{equation}
The twisted superpotential (\ref{2df4d}) serves as the Yang-Yang function of some quantum integrable system. It was further observed in \cite{Nekrasov:2009rc} that the parameter $\varepsilon$ is the Planck constant and can be zero in the classical limit. In this regard, the prepotential ${\cal F} \left(\sigma; q\right)$ can be used to describe the classical integrable system \cite{Gorsky:1995zq, Donagi:1995cf}.

All of these models have a similar structure as in Eq.(\ref{EKSC}), which indicates that there is an emergent spin chain in low-energy physics. However, the full soliton spectrum has not been investigated in the literature, which we leave to future work\cite{Gu:20xx}.\\

\noindent{\underline{Gukov-Witten Surface Operators}}\\

 Gukov-Witten surface operators, like Wilson and 't Hooft operators, can be used to probe the theory, label the phases, etc. They have been defined and investigated in \cite{Gukov:2006jk,Gukov:2008sn}. See also a review in \cite{Gukov:2014gja}. One way to define a surface operator in quantum field theory formalized via a Feynman path integral: is as singularities or boundary conditions for the gauge field $A_{\mu}$ as well as other fields in a vector multiplet along a surface $D$ in a higher-dimensional space-time \cite{Kapustin:2005py}.

 In a four-dimensional ${\cal N}$=2 supersymmetric gauge theory, the $\Omega$ deformed space-time, $\mathbb{R}^{2}_{\widetilde{\varepsilon}}\times \mathbb{R}^{2}_{\varepsilon}$, has a natural explanation in a gauge theory with surface operators. It suggests that the twisted superpotential $\widetilde{W}_{{\rm eff}}$ is the same as the one obtained from Eq.(\ref{2df4d}). However, a surface operator has more information about the dynamics: one can define a UV GLSM on $D$ that includes domain walls. And the twisted superpotential can be replaced by the superpotential for the (non)-abelian mirror.

In a three-dimensional ${\cal N}$=2 Chern-Simons matter theory on $S_{t}^{1}\times_{\mathfrak{q}}D$ with a torus boundary $S_{ t}^{1}\times S_{ Sp}=\partial\left( S_{t}^{1}\times_{\mathfrak{q}}D \right)$.  Assuming the radii of the ``temporal" circle, $S_{t}^{1}$, and ``spatial" circle, $S_{ Sp}$, of the boundary torus, are $R_{1}$ and $R_{2}$, respectively, then define a parameter
\begin{equation}\label{}
  \mathfrak{q}=e^{\frac{R_{1}}{R_{2}}}.
\end{equation}
The surface operator is also useful in this case and supported on $D$. The twisted effective superpotential can be extracted from the vortex partition function on space-time $S_{Sp}$ by
\begin{equation}\label{}
  Z_{{\rm vortex}}\simeq\exp\left(\frac{\widetilde{W}_{{\rm eff}}}{\log \mathfrak{q}}+\ldots\right)
\end{equation}
in the limit, where $\mathfrak{q}\mapsto 1$. See \cite{Gadde:2013wq} for more about surface operators in the 3d theory and their connections to integrable systems. See also \cite{Gukov:2014gja} for surface operators in other-dimensional quantum field theory.

So far, we have reviewed several kinds of two-dimensional supersymmetric gauge theories with four supercharges. In the next section, we will show how a Heisenberg spin chain emerges from the two-dimensional supersymmetric gauge theory.

\section{Heisenberg Spin Chain}\label{Sec:3}
In this section, we introduce how the Heisenberg spin chain appears at an intermediate scale of a gauge theory.\\

\noindent{\underline{Why ${\cal N}$=(2,2)?}} \\

In section \ref{Sec:2.1.3}, we have argued why two dimensions are crucial for obtaining an integrable system from a quantum field theory with infinite degrees of freedom. Now, we will show why ${\cal N}$=(2,2) is necessary for constructing a Heisenberg spin chain. In the spin chain, the fundamental physical operators are spin operators such as $S_{\pm,i}$ and $S_{z,i}$. They obey an algebraic relation:
\begin{equation}\label{}
  [S_{+,i},S_{-,i}]=2S_{z,i}.
\end{equation}
In examples such as $U(k)$ gauge theories, those operators can be constructed from emergent fermions $\lambda_{\pm,i}$ and ${\bar\lambda}_{\pm,i}$ as:
\begin{equation}\label{}
  S_{+,i}={\bar \lambda}_{+,i}\lambda_{-,i},\quad\quad S_{-,i}={\bar \lambda}_{-,i}\lambda_{+,i},\quad\quad S_{z,i}=\frac{1}{2}\left({\bar \lambda}_{+,i}\lambda_{+,i}-{\bar \lambda}_{-,i}\lambda_{-,i}\right).
\end{equation}
 Each site has a two-dimensional vector space. However, in GLSM for $T^{\ast}Gr(k;N)$ discussed in section \ref{Sec:2.1.2}. The $\mathbb{Z}_{2s}$ global symmetry induces an internal index $a\in\{1,\cdots,2s\}$ in fermions which suggests the following replacement: ${\bar \lambda}_{\pm,i}\mapsto\sum_{a}{\bar \lambda}^{a}_{\pm,i}$ and $\lambda_{\pm,i}\mapsto\sum_{a}\lambda^{a}_{\pm,i}$. In this case, the spin operators generate the spin-$s$ representations of the $SU(2)$ group. On the other hand, in a quiver gauge theory, we write the emergent fermions as $\lambda^{n}_{\pm,i}$ and ${\bar\lambda}^{n}_{\pm,i}$, where the index $n$ and $\pm$ characterize jointly for a finite subgroup of $SU(2)$, which is associated with the $ADE$ classification of a quiver gauge theory. In this situation, the spin operators have one more index. Thus, one could construct more states, which are representations of an $ADE$ group for a quiver gauge theory. In fact, the spin-$s$ $SU(2)$ XXX model can alternatively be written as a $SU(2s+1)$-quiver XXX model.

  Focus on $U(k)$ gauge theories, one can easily observe that the operator $S_{z}=\sum_{i}S_{z,i}$ is related to the axial R-symmetry operator $F_{A}$ by
\begin{equation}\label{}
  S_{z}:=\frac{1}{2}F_{A}.
\end{equation}
In section \ref{Sec:2.1.3}, we have already shown that the vacuum configuration of the supersymmetric gauge theory can be expressed as a circle chain on the Coulomb branch. As we have seen in Eq.(\ref{FDM}), these emergent fermions can be constructed from the vector multiplets of the ${\cal N}$=(2,2) supersymmetric gauge theories in a long route: RG-flow, changing variables, integrating over $\sigma$-fields, etc. Eventually, one has
\begin{equation}\label{}\nonumber
   {\bar \lambda}_{\pm,j}=:\int d\widehat{{\bar\sigma}}d\widehat{\sigma} \delta^{2}\left(\widehat{\sigma}-\widehat{\sigma}_{j}\right){\bar \lambda}_{\pm}, \quad { \lambda}_{\pm,j}=: \int d\widehat{{\bar\sigma}}d\widehat{\sigma} \delta^{2}\left(\widehat{\sigma}-\widehat{\sigma}_{j}\right)\lambda_{\pm}.
\end{equation}
Based on the above dictionary, we expect that the spin operators of the Heisenberg spin chain can naturally emerge from a 2d ${\cal N}$=(2,2) gauge theory. However, we want to point out that not every operator in a spin chain can have a direct correspondence in a single gauge theory: the operators $S_{+}$ and $S_{-}$ change the rank of a gauge group in the dictionary. Moreover, they do not commute with $S_{z}$, a generator for $U(1)_{A}$ global symmetry in one gauge theory. However, this does not say these operators are useless in constructing a spin chain for one gauge theory. In fact, we will see that the Hamiltonian built from these operators does commute with all global symmetries. This phenomenon indicates an interrelation among gauge theories from the perspective of the spin chain. We will see more about this in section \ref{Sec:4}.

 Before going next, we want to mention that spin chain operators can also be constructed from bosonic variables \cite{Faddeev:1996iy}. Although we still did not find a straightforward meaning for these bosonic variables in gauge theory, we expect the Boson–Fermion correspondence may shed light on this. \\

\noindent{\underline{Domain Walls}}\\

The domain walls in supersymmetric gauge theory can also be represented by spin operators. Since the domain wall with one electronic charge, $\Phi$, interpolates two adjacent vacua defining a map
\begin{equation}\label{}
 \phi: \quad \ldots{\bar \lambda}_{-, l}{\bar \lambda}_{+, l+1}\ldots \mapsto \ldots{\bar \lambda}_{+, l} {\bar \lambda}_{-, l+1}\ldots.
\end{equation}
The amplitude of this is, of course, non-vanishing at the intermediate scale of a supersymmetric gauge theory, which is approximately equal to
\begin{equation}\label{}
  e^{-\frac{M_{ l( l+1)}}{\mu}}.
\end{equation}
This map, $\phi$, can actually be represented, in the spin chain, by a composite operator
\begin{equation}\label{}
  \phi:=J_{1,l}{\cal D}_{l}= J_{1,l}S_{-, l+1}S_{+,l} \quad {\rm for} \ l\in\{1,\cdots,N-1\},
\end{equation}
where $J_{1,l}$ is a parameter that measures the amplitude of this domain wall, and its detailed expression may not be crucial for our purposes in this paper. In periodic $N$-sites, we have
\begin{equation}\label{}
  {\cal D}_{N}= S_{-, 1}S_{+,N}.
\end{equation}
While for a twisted boundary condition:
 \begin{equation}\label{}
  {\cal D}_{N}= \widetilde{q}S_{-, 1}S_{+,N},
\end{equation}
where $\widetilde{q}=e^{-\widetilde{t}}$. The anti-domain wall is also straightforward
\begin{equation}\label{}
  {\bar\phi}:= J_{1,l}{\bar {\cal D}}_{l}=J_{1,l}S_{+, l+1}S_{-,l}.
\end{equation}
The higher gauge charge domain walls follow a similar construction too. The amplitude of a charge-$\ell$ domain wall, $\wedge^{\ell}\Phi$, is proportional to
\begin{equation}\label{}
   e^{-\frac{M_{ l( l+\ell)}}{\mu}}.
\end{equation}
These can also be expressed in terms of the spin operators
\begin{equation}\label{}
  J_{\ell,l}{\cal D}_{[l}\cdots{\cal D}_{l+\ell-1]}.
\end{equation}
The relation $J_{\ell,l}=\left(J_{1,l}\right)^{\ell}$ is not guaranteed in general. Here, we list some short examples. If $\ell=2$, we have
\begin{equation}\label{}
  {\cal D}_{[l}{\cal D}_{l+1]}=-S_{z,l+1}S_{+,l}S_{-,l+2},
\end{equation}
and if $\ell=2$, the formula is
\begin{equation}\label{}
  {\cal D}_{[l}{\cal D}_{l+1}{\cal D}_{l+2]}=-\frac{1}{3}S_{-,l+3}S_{+,l}\left(S_{-,l+1}S_{+,l+1}S_{z,l+2}+S_{+,l+2}S_{-,l+2}S_{z,l+1}\right).
\end{equation}

\noindent{\underline{The Affine Nil-Temperley–Lieb Algebra}}\\

In the above computation, we have used the affine nil-Temperley-Lieb algebra, which can be proved by a direct computation from our definition
\begin{equation}\label{}
  \left({\cal D}_{i}\right)^{2}={\cal D}_{i}{\cal D}_{i+1}{\cal D}_{i}={\cal D}_{i+1}{\cal D}_{i}{\cal D}_{i+1}=0,\quad\quad {\cal D}_{i}{\cal D}_{j}={\cal D}_{j}{\cal D}_{i}\quad {\rm if}\ i-j\neq \pm 1\ {\rm mod}\ N
\end{equation}
where all indices are defined modulo $N$. So the affine nil-Temperley-Lieb algebra is a consequence of the dynamics of the (anti-)domain walls\\

\noindent{\underline{(In-)homogeneous Spin Chain}}\\

It is easy to see that $J_{1,l}=J_{1, j}$ if $M_{ l( l+1)}=M_{ j( j+1)}$, so the generic twisted masses correspond to an inhomogeneous spin chain. In a two-dimensional gauge theory, there is a so-called decoupling limit of matters by sending some mass parameters to infinity. For example, if we take the mass $m_{i}\mapsto\infty$, the matter $\Phi_{i}$ with this mass will not affect the dynamics of the left degrees of freedom. Let us use $\mathbb{P}^{N}$ as an example. If we turn on the mass $m_{N+1}$ for the $\Phi_{N+1}$ field, the vacuum equation is
\begin{equation}\label{}
  \sigma^{N}\left(\sigma+m_{N+1}\right)=\left(\widetilde{\Lambda}_{N+1}\right)^{N+1}.
\end{equation}
Now we define a new dynamical scale in the limit where $m_{N+1}$ goes to infinity:
\begin{equation}\label{}
  \left(\Lambda_{N}\right)^{N}=\lim_{m_{N+1}\mapsto\infty}\frac{\left(\widetilde{\Lambda}_{N+1}\right)^{N+1}}{m_{N+1}}.
\end{equation}
Then the left equation is
\begin{equation}\label{}
  \sigma^{N}=  \left(\Lambda_{N}\right)^{N},
\end{equation}
which is the vacuum equation for $\mathbb{P}^{N-1}$. It also corresponds to a spin chain. So the decoupling limit, by sending $k$ of $m_{i}$ to $\infty$, induces a map between the spin chains with different lengths:
\begin{equation}\label{}
 {\rm Spin\ chain\ length}\ N \mapsto  {\rm Spin\ chain\ length}\ N-k.
\end{equation}\\

\noindent{\underline{Symmetries}}\\

Symmetries of the two-dimensional gauge theory can be also reflected in the spin chain system. The center group $\mathbb{Z}_{N}$ of the flavor symmetry corresponds to the translation symmetry of the $N$-sites spin chain. It was first observed in \cite{Nekrasov:2009uh} that the rank of gauge group $k$ is the number of magnons on the spin chain. It is natural to expect that our previous definition
\begin{equation}\label{TMM}
  e_{k}\left(\sigma_{a}\right)=e^{i\sum^{k}_{a=1}p_{a}} \quad{\rm or} \quad e_{k}\left(\frac{\sigma_{a}+is}{\sigma_{a}-is}\right)=e^{i\sum^{k}_{a=1}p_{a}},
\end{equation}
as a gauge-invariant physical operator of the Weyl group, is a group representation of the center symmetry. The notation $p_{a}$ can be understood as the momentum of the $a$-th magnon if it is real. For example, it can happen when $\left(e_{k}\right)^{N}=1$. So one can easily see that $\sum^{k}_{a=1}p_{a}$ is the total momentum of the $k$ magnons. There are other finite symmetries as well, such as ${\cal P}$, ${\cal T}$, and ${\cal C}$, in a quantum field theory that act on the field variables, for $1\leq i\leq N$, as
\begin{eqnarray}
  {\cal P}{\bar \lambda}_{\pm, i}{\cal P}^{-1}  = {\bar \lambda}_{\mp, i}\quad&&\quad {\cal P} \lambda_{\pm, i}{\cal P}^{-1}  = \lambda_{\mp, i}, \\\nonumber
  {\cal T}{\bar \lambda}_{\pm, i}{\cal T}^{-1} =  \lambda_{\mp, i}\quad&&\quad  {\cal T}\lambda_{\pm, i}{\cal T}^{-1} =  {\bar\lambda}_{\mp, i}, \\\nonumber
  {\cal C}{\bar \lambda}_{\pm, i}{\cal C}^{-1} = \lambda_{\pm, i}\quad&&\quad {\cal C} \lambda_{\pm, i}{\cal C}^{-1}={\bar \lambda}_{\pm, i}.
\end{eqnarray}

The associated spin operators will be transformed as
\begin{eqnarray}
  {\cal P}S_{\pm, i}{\cal P}^{-1}  = S_{\mp, i} \quad&&\quad  {\cal P}S_{z, i}{\cal P}^{-1}  = -S_{z, i} ,\\\nonumber
    {\cal T}S_{\pm, i}{\cal T}^{-1} = - S_{\pm, i}\quad&&\quad {\cal T}S_{z, i}{\cal T}^{-1}  = S_{z, i}, \\\nonumber
  {\cal C}S_{\pm, i}{\cal C}^{-1} =-S_{\mp, i}    \quad&&\quad {\cal C}S_{z, i}{\cal C}^{-1} =-S_{z, i} .
\end{eqnarray}
These symmetries act on the domain walls as
\begin{eqnarray}
  {\cal P}{\cal D}_{ i}{\cal P}^{-1}  = {\bar{\cal D}}_{ i} \quad&&\quad  {\cal P}{\cal D}_{ N}{\cal P}^{-1}  = \left(\frac{\widetilde{q}}{|\widetilde{q}|}\right)^{2}{\bar{\cal D}}_{ N} ,\\\nonumber
    {\cal T}{\cal D}_{ i}{\cal T}^{-1} = {\cal D}_{ i}\quad&&\quad {\cal T}{\cal D}_{ N}{\cal T}^{-1}  = \left(\frac{\widetilde{q}}{|\widetilde{q}|}\right)^{-2}{\cal D}_{ N}, \\\nonumber
  {\cal C}{\cal D}_{ i}{\cal C}^{-1} ={\bar{\cal D}}_{ i}   \quad&&\quad {\cal C}{\cal D}_{ N}{\cal C}^{-1} ={\bar{\cal D}}_{ N} ,
\end{eqnarray}
where we have used the fact that $ {\cal T}i{\cal T}^{-1}=-i$ for $i^{2}+1=0$. From the above, one can observe that ${\cal P}$ and ${\cal T}$ may be violated individually unless $\widetilde{q}=\pm1$. However, we will see later that the scattering factor could also break the ${\cal T}$-symmetry if it is not a pure phase factor. On the other hand, ${\cal CPT}$ symmetry is preserved as expected. The authors in \cite{Korff:2010} also point out a finite symmetry of the spin chain by mapping the index of sites as $i\mapsto N+1-i$, which is useful for understanding the operation of Hermitian conjugate of composite operators.\\

\noindent{\underline{Closed Versus Open Spin Chain}}\\

$\bullet$ When the scattering factor, $S_{ab}$, is a pure phase: Besides a usual closed spin chain with $\widetilde{q}=1$ defined in the literature, we claim that the anti-periodic spin chain with $\widetilde{q}=-1$ is also a \textit{closed} one. On the other hand, the open spin chain in this situation has $\widetilde{q}\neq\pm 1$.

$\bullet$ When the scattering factor is not a pure phase: It is always an open spin chain for any $\widetilde{q}$.\\

\noindent{\underline{Hamiltonians}}\\

Eq.(\ref{TMM}) is a representation of the center symmetry $\mathbb{Z}_{N}$ of the flavor group. So one can expect that the number of independent conserved charges is $N$. Let us denote them as $h_{i}$'s, which satisfy the commutative relations
 \begin{equation}\label{}
   \left[h_{i},h_{j}\right]=0 \quad\quad {\rm for} \quad i\neq j.
 \end{equation}
These generators can also be understood as
\begin{equation}\label{}
  e^{i\cdot h_{j}}e^{i\sum^{k}_{a=1}p_{a}}e^{-i\cdot h_{j}}:=e^{i\left(\sum^{k}_{a=1}p_{a}+j\right)}.
\end{equation}
One may propose that $h_{1}$ corresponds to the fundamental Hamiltonian $H$ of the spin chain because it generates the shift of the total momentum by one unit: $\sum^{k}_{a=1}p_{a}\mapsto \sum^{k}_{a=1}p_{a}+1$. Since there could be no time-reversal symmetry in a general situation. One may define, up to an overall constant, the fundamental complex Hamiltonian as
\begin{equation}\label{HFH}
  h_{1}:=\sum^{N}_{i=1}{\cal D}_{i}.
\end{equation}\\
However, the above complex Hamiltonian can not capture the interaction between the ``complex magnons". This issue can be solved by adding an ``interaction" term to the Hamiltonian as:
 \begin{equation}\label{}
   h_{1}=\sum^{N}_{i=1}\left({\cal D}_{i}+f({\cal D}_{i})\right).
 \end{equation}
The complex conjugate hamiltonian can be obtained by the symmetry: ${\bar h}_{1}={\cal C}h_{1}{\cal C}^{-1}$. So we can only focus on the holomorphic Hamiltonian.

If the scattering factor is a pure phase and $\widetilde{q}=\pm1$, then the system is a closed spin chain that has two more finite symmetries: ${\cal T}$ and ${\cal P}$. Therefore, the Hamiltonian is a hermitian operator. There is an easy fix by changing the Hamiltonian in (\ref{HFH}) to be
\begin{equation}\label{FTOH}
  H:=h_{1}+{\bar h}_{1}=\sum^{N}_{i=1}\left({\cal D}_{i}+{\bar{\cal D}}_{i}\right).
\end{equation}
 The conjugation part, ${\bar h}_{1}$, induces a minus one-unit shift to the total momentum by \\       $\sum^{k}_{a=1}p_{a}\mapsto\sum^{k}_{a=1}p_{a}-1$. This operation may look odd. However, since $h_{1}$ commutes with ${\bar h}_{1}$, it is well-defined to take a linear combination of them. In fact, $H$ in (\ref{FTOH}) is the Hamiltonian of the so-called XX spin chain that describes the \textit{free fermion} system.

 To include the dynamical information of magnons. We need to further improve the Hamiltonian as
 \begin{equation}\label{}
   H=\sum^{N}_{i=1}\left({\cal D}_{i}+{\bar{\cal D}}_{i}\right)+f({\cal D}_{i}, {\bar{\cal D}}_{i}),
 \end{equation}
 where $f({\cal D}_{i}, {\bar{\cal D}}_{i})$ commutates with all finite symmetries. The detailed expression of $f({\cal D}_{i}, {\bar{\cal D}}_{i})$ can be fixed if we further impose a condition that the ground state wave functions $|\omega\rangle_{k}$ are eigenstates of the Hamiltonian:
 \begin{equation}\label{}
   H|\omega\rangle_{k}:=E_{k}|\omega\rangle_{k}.
 \end{equation}
 We will show how to derive $f$ for some cases in the following sections. The operator $\sum^{N}_{i=1}\left({\cal D}_{i}+{\bar{\cal D}}_{i}\right)$ commutes with the operator $\sum^{N}_{i=1}S_{z,i}$ already. Thus, if the operator $f({\cal D}_{i}, {\bar{\cal D}}_{i})$ also commutes with the operator $\sum^{N}_{i=1}S_{z,i}$, we may consider a linear combination of these operators to give a modified Hamiltonian as:
 \begin{equation}\label{}
   H_{B}=\sum^{N}_{i=1}\left({\cal D}_{i}+{\bar{\cal D}}_{i}\right)+f({\cal D}_{i}, {\bar{\cal D}}_{i})+B\sum^{N}_{i=1}S_{z,i},
 \end{equation}
where $B$ is the ``background magnetic field" in the spin chain system. This new Hamiltonian breaks the ${\cal C}$ and ${\cal P}$ symmetries explicitly. So it would be interesting to understand what the $B$ field corresponds to in a field theory.

Before going to concrete examples, we want to comment on the relationship between the matter fields in a gauge theory and the Hamiltonian in the corresponding spin chain. Notice that the fundamental field $\Phi$ or $\widetilde{\Phi}^{\dag}$ ($\widetilde{\Phi}$ is the anti-fundamental representation) is the domain wall. They correspond to the operator $\sum^{N}_{i=1}{\cal D}_{i}$ in the spin chain. Similarly, the conjugation field ${\bar\Phi}$ or $\widetilde{\Phi}$ is the anti-domain wall that can be expressed in terms of the spin operators as $\sum^{N}_{i=1}{\bar{\cal D}}_{i}$. There could be other matters in the gauge theory as well. However, for the flavor symmetry to be a simply-connected compact Lie group. Every (finite-dimensional, continuous, complex) irreducible representation of the flavor group is a sub-representation of a tensor product of copies of the fundamental representation, and the anti-fundamental one\footnote{For a general group, this conclusion does not apply.}. Thus, it is natural to expect that their correspondences in the spin chain are composite spin operators: $f({\cal D}_{i}, {\bar{\cal D}}_{i})$.  For example, the adjoint matter $X$ can be expressed in terms of a tensor product between the fundamental and anti-fundamental field by
 \begin{equation}\label{}
   X:=\Phi\otimes {\bar\Phi}-\textbf{I},
 \end{equation}
 where $\textbf{I}$ is an identity operator and can be expressed by
 \begin{multline*}
    \textbf{I}:=\frac{1}{N}\sum^{N}_{i=1}\left(S_{+,i}S_{+,i+1}S_{-,i}S_{-,i+1}+S_{-,i}S_{-,i+1}S_{+,i}S_{+,i+1}\right. \\\nonumber
   \left.+S_{-,i}S_{+,i+1}S_{+,i}S_{-,i+1}+S_{+,i}S_{-,i+1}S_{-,i}S_{+,i+1}\right).
 \end{multline*}

To express $\Phi\otimes {\bar\Phi}$ in terms of spin operators, we first notice that the adjoint field is a gauge-neutral field that can only define on the vacuum configuration where $\sigma\left(-\infty\right)=\sigma\left(+\infty\right)$. Furthermore, since the adjoint field commutes with the center group $\mathbb{Z}_{N}$, we propose the following map
\begin{equation}\label{}
  \Phi\otimes {\bar\Phi}:=\frac{1}{2}\sum^{N}_{i=1}\left({\cal D}_{i}{\bar{\cal D}}_{i}+{\bar{\cal D}}_{i}{\cal D}_{i}\right),
\end{equation}
where we have applied the Weyl ordering to the right-hand side of the above equation. A direct computation shows that
\begin{equation}\label{}
  \left(\frac{N}{4}-1\right)\textbf{I}-X:=\sum^{N}_{i=1}S_{z,i}S_{z,i+1}.
\end{equation}
 The gauge field is also an adjoint representation. However, its roles are different. It constrains the dynamics of the system as expressed in terms of spin operators:
\begin{equation}\label{}
 \left( S_{+,i}\right)^{2}=\left( S_{-,i}\right)^{2}=0.
\end{equation}
 Besides the above, the left Weyl group gauge symmetry imposes an ordering of the spin chain excitations: magnons. The dictionary for a general representation can be studied similarly.\\

\noindent{\underline{Lattice Supersymmetry}} \\

Our construction of Hamiltonian relies on the (anti)-domain walls configuration, i.e, BPS solitons. The BPS bound is saturated when half (in our situation) of the SUSY generators are unbroken, which indicates that the spin chain has hidden supersymmetry. However, the Hamiltonian itself does not see it explicitly. Finally, we want to mention that there have been some studies of the so-called lattice supersymmetry in the literature already \cite{Fendley:2002sg,Fendley:2003je}. Our paper not only provides an underlying physical explanation for the Lattice supersymmetry but also offers new insight into constructing supersymmetric algebra for spin chain models. However, we will not focus on any technical discussion for this in this paper. And in the following several subsections, we will focus on concrete examples by applying the dictionary we found.

 \subsection{Examples: XX Model And ${\rm XXX_{s}}$ Model}
 In this section, we will study examples of two-dimensional gauge theories and their relations with the Heisenberg spin chain. Before that, we should mention that it is known that the Gromov-Witten invariants can be understood and computed from GLSMs and their mirrors \cite{Gu:2020ana}. On the other hand, mathematicians \cite{Korff:2010} have also constructed five-vertex integrable models to prove statements for Gromov-Witten theory of  $Gr(k;N)$. The physical reason for the ``three faces of one thing" has been explained in the previous section. In this section, we make this more transparent by studying examples.

 \subsubsection{$Gr(k;N)$ As The XX Model}\label{GRXX}
 The vacuum equations of Grassmannian $Gr(k;N)$ are
 \begin{equation}\label{}
   \left(\widehat{\sigma}_{a}\right)^{N}=1,
 \end{equation}
 where $\widehat{\sigma}_{a}=\sigma_{a}\Lambda^{-1}(-1)^{\frac{1-k}{N}}$. The solutions of the above equation induce an $N$ sites spin chain on the Coulomb branch field variable $\widehat{\sigma}_{a}$. The ground state wave functions are proposed before (\ref{GRKGSWF}) as
 \begin{equation}
  |\omega\rangle_{k}:= A_{k}\sum^{N}_{1\leq j_{1}<,\cdots,< j_{k}\leq N}\prod^{k}_{a=1}\sum_{{\cal W}\in S_{k}}\left(\widehat{\sigma}_{{\cal W}(a)}\right)^{j_{a}}|j_{1},\cdots ,j_{k}\rangle_{k}.
\end{equation}
Since there is no interacting term in the ground state wave functions above, we propose the Hamiltonian of this spin chain is
\begin{equation}\label{}
  H=\sum^{N}_{i=1} \left({\cal D}_{i}+{\bar{\cal D}}_{i}\right).
\end{equation}
This is the Hamiltonian of the XX model. A straightforward calculation gives that
\begin{equation}\label{}
  H|\omega\rangle_{k}=\left(e_{1}(\widehat{\sigma})+e_{1}(\widehat{{\bar\sigma}})\right)|\omega\rangle_{k},
\end{equation}
where we have used the fact that $\widehat{{\bar\sigma}}=\widehat{\sigma}^{-1}$. We can further consider higher Hamiltonians as
\begin{equation}\label{}
  H_{r}=e_{r}({\cal D}_{i})+e_{r}({\bar{\cal D}}_{i}),
\end{equation}
where the definition of $e_{r}({\cal D}_{i})$ can be found in \cite{Postnikov:2022}. It is
\begin{equation}\label{}
  e_{r}({\cal D}_{i})=\sum_{|I|=r}\prod^{\circlearrowright}_{i\in I}{\cal D}_{i},
\end{equation}
where $I$ is the proper subset with the $r$ number of the overall $N$ sites, and $\prod^{\circlearrowright}_{i\in I}{\cal D}_{i}$ is the product of ${\cal D}_{i}$, $i\in I$, taken in an order such that if $i$, $i+1\in I$ then ${\cal D}_{i+1}$ goes before ${\cal D}_{i}$. Actually, in \cite{Postnikov:2022}, the author further defines the basis
\begin{equation}\label{}
  h_{r}({\cal D}_{i})=\sum_{|I|=r}\prod^{\circlearrowleft}_{i\in I}{\cal D}_{i},
\end{equation}
where the notation $\prod^{\circlearrowleft}_{i\in I}{\cal D}_{i}$ represents the reversing ``cyclic order" of ${\cal D}_{i}'s$ in $\prod^{\circlearrowright}_{i\in I}{\cal D}_{i}$. For example,
\begin{eqnarray}
  e_{2}\left({\cal D}_{i}\right) &=& {\cal D}_{2}{\cal D}_{1}+{\cal D}_{3}{\cal D}_{2}+\cdots+ {\cal D}_{n}{\cal D}_{n-1}+{\cal D}_{1}{\cal D}_{n}+\sum_{|i-j|>1}{\cal D}_{i}{\cal D}_{j},\\\nonumber
 h_{2}\left({\cal D}_{i}\right) &=& {\cal D}_{1}{\cal D}_{2}+{\cal D}_{2}{\cal D}_{3}+\cdots+ {\cal D}_{n-1}{\cal D}_{n}+{\cal D}_{n}{\cal D}_{1}+\sum_{|i-j|>1}{\cal D}_{i}{\cal D}_{j}.
\end{eqnarray}

A direct computation suggests that
\begin{equation}\label{}
  e_{r}({\cal D}_{i})|\omega\rangle_{k}=e_{r}(\sigma)|\omega\rangle_{k},\quad\quad h_{r}({\cal D}_{i})|\omega\rangle_{k}=h_{r}(\sigma)|\omega\rangle_{k}.
\end{equation}

The Schur operators can also be constructed from $e_{i}$ and $h_{j}$. And when they act on $|\omega\rangle_{k}$, one can find the eigenvalues of these Schur-operators are the Schubert basis, i.e. Schur functions of Chern-roots $\sigma_{a}$, of Gromov-Witten theory of Grassmannian. See \cite{Korff:2010} for more details.

Since we consider the two-dimensional gauge theory at the intermediate scale, the $\mathbb{Z}_{2N}$ axial symmetry is not spontaneously broken to $\mathbb{Z}_{2}$. While this spontaneous breaking of symmetry only happens in the far infrared. Based on our dictionary proposed in section \ref{Sec:3}, the spin operator $S_{z}=\frac{1}{2}F_{A}$ acts on the wave functions $|\omega\rangle_{k}$ to give
\begin{equation}\label{}
  S_{z}|\omega\rangle_{k}=(k-\frac{N}{2})|\omega\rangle_{k}.
\end{equation}
 Since $S_{z}$ commutes with the Hamiltonian in this case, it is a conserved quantity counting the $z$-component spin of all sites. The vector R-symmetry $F_{V}$ has a meaning in the spin chain as well. It counts the number of sites of the spin chain:
 \begin{equation}\label{}
   F_{V}|\omega\rangle_{k}=-N|\omega\rangle_{k}.
 \end{equation}
 So the operator $S_{z}-\frac{1}{2}F_{_{V}}$ counts the number of magnons. It is $k$ for $Gr(k;N)$.

 If we turn on generic twisted masses, the Hamiltonian will be corrected to
 \begin{equation}\label{}
     H=\sum^{N}_{i=1} J_{i}\left({\cal D}_{i}+{\bar{\cal D}}_{i}\right),
 \end{equation}
 where $J_{i}\neq J_{j}$ if $i\neq j$. The decoupling limit is operated by sending some twisted masses to infinity. Some of $J\mapsto 0$ will become zero-value in this limit.

 \subsubsection{$Gr(k;N)$ As An Open Spin Chain}
If we use $\sigma_{a}$ instead of $\widehat{\sigma}_{a}$, and set the physical scale $\mu=1$, the vacuum equations would like to be
\begin{equation}\label{}
  \left(\sigma_{a}\right)^{N}=\widetilde{q}=(-1)^{k-1}q.
\end{equation}
We will then get an open spin chain if $\widetilde{q}\neq\pm1$. The fundamental Hamiltonian is chosen to be
\begin{equation}\label{}
  h:=\sum^{N}_{i=1}{\cal D}_{i}.
\end{equation}
So by focusing only on the holomorphic part, the investigation is in parallel with the previous section.

\subsubsection{$T^{\ast}Gr(k;N)$ As The ${\rm XXX_{s}}$ Model}
The vacuum equations of $T^{\ast}Gr(k;N)$ can be investigated on the Coulomb branch
\begin{equation}\label{VETGRS}
  \left(\frac{\widehat{\sigma}_{a}+is}{\widehat{\sigma}_{a}-is}\right)^{N}=(-1)^{N}q\prod_{b\neq a}\left(\frac{\widehat{\sigma}_{a}-\widehat{\sigma}_{b}+i}{\widehat{\sigma}_{a}-\widehat{\sigma}_{b}-i}\right).
\end{equation}
When $s=\frac{1}{2}$, the ground state wave functions are proposed to be before
\begin{equation}
    A_{k}\sum^{N}_{1\leq j_{1}<\cdots <j_{k}\leq N}\sum_{{\cal W}\in S_{k}}S_{{\cal W} \left(a\right)}e^{\sum^{k}_{a=1}ip_{{\cal W} \left(a\right)}j_{a}}|j_{1}<\cdots<j_{k}\rangle_{k}.
\end{equation}
When $(-1)^{N}q=\pm1$\footnote{Only $(-1)^{N}q=1$ in (\ref{VETGRS}) corresponds to the usual Bethe ansatz equation. When $(-1)^{N}q=-1$, it is an anti-periodic boundary condition. The overall scattering phase factor will be shifted by a $\pi$. However, only one choice, in general, is a $regular$ one. The meaning of regular in our paper is that all the vacua solely locate on the Coulomb branch \cite{Hori:2011pd}. So we keep both.}, it is known that the above wave functions are eigenstates of the hermitian Hamiltonian
\begin{equation}\label{}
  H=\sum^{N}_{i=1}\left({\cal D}_{i}+{\bar{\cal D}_{i}}-{\cal D}_{i}{\bar{\cal D}_{i}}-{\bar{\cal D}_{i}}{\cal D}_{i}+\frac{N\textbf{I}}{2}\right)+c\cdot\textbf{I}.
\end{equation}
Or writing it in terms of the spin operators
\begin{equation}\label{}
   H=\sum^{N}_{i=1}\left(S_{+,i+1}S_{-,i}+S_{-,i+1}S_{+,i}+2S_{z,i+1}S_{z,i}\right)+c\cdot\textbf{I}.
\end{equation}
This is exactly the Hamiltonian of ${\rm XXX_{\frac{1}{2}}}$-model. One can further define higher Hamiltonians by
\begin{equation}\label{}
  H_{r}=e_{r}({\cal D}_{i})+e_{r}({\bar{\cal D}}_{i})+f_{r}\left({\cal D}_{i}, {\bar{\cal D}}_{i}\right).
\end{equation}
The formulation of $f_{r}$ can be derived if we require the wave functions are eigenstates of $H_{r}$. Their relation to the quantum cohomology of the cotangent bundle to the Grassmannian \cite{Maulik:2012wi} will be investigated in \cite{Gu:20xx}. The construction for a general ${\rm XXX_{s>\frac{1}{2}}}$-model is similar.\\

\noindent{\underline{Symmetry}}\\

The XX model has a $U(1)$ symmetry generated by the rotation of the $z$-axis. However, the XXX model has an enlarged one. To see this, we rewrite the Hamiltonian of XXX model as
 \begin{equation}\label{}
   H=\sum^{N}_{i=1}2\left(S_{x,i+1}S_{x,i}+S_{y,i+1}S_{y,i}+S_{z,i+1}S_{z,i}\right)+c\cdot\textbf{I},
\end{equation}
where we have used $S_{\pm}=S_{x}\pm iS_{y}$. One can easily observe that the Hamiltonian of ${\rm XXX_{s}}$ model has a manifestly $SU(2)$ symmetry with the Lie algebra $\mathfrak{su}(2)$ to be
\begin{equation}\label{}
 {\cal R}_{\alpha}=\frac{1}{2}\sum^{N}_{i=1}S_{\alpha,i},\quad{\rm where} \quad \alpha=x,y,z.
\end{equation}
We know that the spin operator ${\cal R}_{z}$ is the generator of the axial $U(1)$ R-symmetry in the ${\cal N}=(2,2)$ supersymmetric gauge theory. So only the diagonal part of the $SU(2)$ symmetry can be seen directly from the UV supersymmetric gauge theory. However, we should notice that the low energy limit of the gauge theory is a sigma model on the cotangent bundle to the Grassmannian. It is a hyperK\"{a}hler manifold, as required by the ${\cal N}=(4,4)$ supersymmetry in two dimensions. Then, indeed an emergent $SU(2)$ axial R-symmetry appears in low-energy physics. Thus, we claim that this $SU(2)$ axial R-symmetry is the origin of the $SU(2)$ symmetry of ${\rm XXX_{s}}$ model.

 \subsection{K-theoretic XX Model And ${\rm XXZ_{s}}$ Model}
The three-dimensional gauge theory has an extra parameter called the Chern-Simons level. The low energy physics, defined on the Coulomb branch on space-time $\mathbb{R}^{2}\times \mathbb{S}^{1}$, depends on the choice of the Chern-Simons level. Thus, the spin chain structure relies on the Chern-Simons level as well.

\subsubsection{K-theoretic XX Model}
If we choose the bare Chern-Simons level $k_{U(1)}=k_{SU(k)}=-\frac{N}{2}$ for the 3d CS-matter theory for Grassmannian $Gr(k;N)$, the vacuum equations are
\begin{equation}\label{}
  \left(1-x_{a}\right)^{N}=(-1)^{k-1}q.
\end{equation}
They are the same as the vacuum equations in two dimensions by replacing $1-x_{a}$ with $\sigma_{a}$. Therefore, the Heisenberg spin chain for this gauge theory is exactly an XX model.

If we pick the bare Chern-Simons level as $k_{U(1)}=-\frac{N}{2}$ and $k_{SU(k)}=k-\frac{N}{2}$, then the vacuum equations are
\begin{equation}\label{}
   \left(1-x_{a}\right)^{N}=q\prod_{b\neq a}\left(-\frac{x_{a}}{x_{b}}\right),
\end{equation}
where the scattering factor is
\begin{equation}\label{}
  S_{ab}=-\frac{x_{a}}{x_{b}}.
\end{equation}
Since the above factor is not a pure phase, no time-reversal symmetry exists for this system. Thus, the fundamental Hamiltonian of this system is a complex one. And its expression was first proposed in \cite{Gorbounov:2014} in a slightly different context:
\begin{eqnarray}\label{GLQKH}
   h=\sum^{N}_{i}{\cal D}_{i}-\sum^{N}_{\mid i_{1}-i_{2}\mid\ {\rm mod}\ N>1}{\cal D}_{i_{1}}{\cal D}_{i_{2}}
   +\sum^{N}_{\mid i_{a}-i_{b}\mid\ {\rm mod}\ N>1}{\cal D}_{i_{1}}{\cal D}_{i_{2}}{\cal D}_{i_{3}}+\cdots.
\end{eqnarray}
The number of sites is $N$, so only finitely many terms act non-trivially, and the series therefore terminates. Let us check whether the ground state wave functions proposed in our paper for this case are indeed the eigenstates of the Hamiltonian (\ref{GLQKH}). We start from the first nontrivial example: $Gr(2;N)$. To shorten the notation, we denote the ground state as $\mid\omega\rangle_{2}=\sum_{1\leq j_{1}<j_{2}\leq N}a(j_{1},j_{2})|j_{1}<j_{2}\rangle_{2}$, where
\begin{equation}\label{}
  a(j_{1},j_{2})=A\left(e^{i\left(p_{1}j_{1}+p_{2}j_{2}\right)}+S_{21}e^{i\left(p_{2}j_{1}+p_{1}j_{2}\right)}\right).
\end{equation}
The boundary condition is $a(j_{1},j_{2}+N)=q\cdot a(j_{2},j_{1})$, which gives
\begin{equation}\label{}
  e^{i\left(p_{1}j_{1}+p_{2}j_{2}\right)}e^{ip_{2}N}+S_{21}e^{ip_{1}N}e^{i\left(p_{2}j_{1}+p_{1}j_{2}\right)}=q\left(e^{i\left(p_{1}j_{2}+p_{2}j_{1}\right)}+S_{21}e^{i\left(p_{2}j_{2}+p_{1}j_{1}\right)}\right).
\end{equation}
The vacuum equations can be read from the above as
\begin{equation}\label{}
  e^{ip_{1}N}=qS_{12},\quad\quad  e^{ip_{2}N}=qS_{21}.
\end{equation}
 So it is consistent. To compute $H\mid\omega\rangle_{2}$, special care is needed when two overturned spins are sitting next to each other. We find
\begin{eqnarray}\label{HGR24}\nonumber
  H\mid\omega\rangle_{2} &=& \sum_{1\leq j_{1}<j_{2}\leq N}(a(j_{1}+1,j_{2})+a(j_{1},j_{2}+1)-a(j_{1}+1,j_{2}+1)) \mid j_{1}<j_{2}\rangle_{2}  \\
   && -\sum_{1\leq j\leq N}\left(a(j+1,j+1)-a(j+1,j+2)\right)\mid j<j+1\rangle_{2}.
\end{eqnarray}

In order to obey the eigenstate condition, the contact terms in the last line of the above equation should be vanishing:
\begin{equation}\label{VCT}
  a(j+1,j+1)-a(j+1,j+2)=0.
\end{equation}
If we test the coefficient as $a(j_{1},j_{2})=Ce^{i\left(p_{1}j_{1}+p_{2}j_{2}\right)}+De^{i\left(p_{1}j_{2}+p_{2}j_{1}\right)}$,
the vanishing contact terms all give
\begin{equation}\label{}
  \frac{C}{D}=-\frac{x_{1}}{x_{2}}.
\end{equation}
This is certainly consistent with our scattering factor $S_{12}$ in the vacuum equations. The procedure applies to a general Grassmannian.

The Hamiltonian actually has a geometrical meaning \cite{Buch:2008}. The eigenvalue of this Hamiltonian is the first Schubert class of the $Gr(k;N)$:
\begin{equation}\label{}
  H\mid \omega\rangle_{k}={\cal O}_{\square}\mid \omega\rangle_{k}
\end{equation}
where
\begin{equation}\label{}
  {\cal O}_{\square}:=1-\prod^{k}_{a=1}x_{a}.
\end{equation}
For example, if $k$=2 the factor in Eq.(\ref{HGR24})
\begin{equation}\label{}
  a(j_{1}+1,j_{2})+a(j_{1},j_{2}+1)-a(j_{1}+1,j_{2}+1)=(z_{1}+z_{2}-z_{1}z_{2})a(j_{1},j_{2})
\end{equation}
where the coefficient $(z_{1}+z_{2}-z_{1}z_{2})=1-x_{1}x_{2}$ is indeed the first Schubert class of $Gr(2,4)$. Thus, one may naturally expect that higher Schubert classes are the eigenvalues of the higher Hamiltonian as well. Quantum K-theory of $Gr(k;N)$ from the integrable model has been investigated in \cite{Gorbounov:2014}, although their construction was based on a five-vertex model rather than a spin chain. We will discuss the connection between these two different approaches in section \ref{Sec:5}.

Finally, we end this section by pointing out that one can still use Schur bundles to study the quantum K-theory of $Gr(k;N)$ mathematically \cite{Gu:2020zpg,Gu:2022yvj}, which may suggest that the above spin chain could be embedded into an inhomogeneous XX model. We leave it for future work.

\subsubsection{K-theoretic XXX Spin Chain: XXZ Model}
The vacuum equations of 3d K-theoretic $T^{\ast}Gr(k;N)$ have been studied in section \ref{KTGR}
\begin{equation}
  \left(\frac{\sin\left(\pi \widetilde{R}\left(\widehat{\sigma}_{a}+\frac{is}{2}\right)\right)}{\sin\left(\pi \widetilde{R}\left(\widehat{\sigma}_{a}-\frac{is}{2}\right)\right)}\right)^{N}=\widetilde{q}\prod^{k}_{b\neq a}\left(\frac{\sin\left(\pi \widetilde{R}\left(\widehat{\sigma}_{a}-\widehat{\sigma}_{b}+i\right)\right)}{\sin\left(\pi \widetilde{R}\left(\widehat{\sigma}_{a}-\widehat{\sigma}_{b}-i\right)\right)}\right),
\end{equation}
where $\widetilde{R}=2Ru$ and $s=\frac{1}{2}$ or 1. Under the rescaling symmetry of $\mathbb{R}^{2}$: $u\mapsto \mathfrak{r} u$, the field variables change as
\begin{equation}\label{}
  \sigma_{a}\mapsto \mathfrak{r}\sigma_{a}.
\end{equation}
So $\widehat{\sigma}_{a}$ is rescaling invariant. It is obvious that the finite radius of $\mathbb{S}^{1}$, $R$, is invariant under the rescaling of $\mathbb{R}^{2}$, so $\widetilde{R}$ is not rescaling invariant. Thus, the vacuum equations are not describing a conformal field theory. This says that the K-theoretic XXX model only has the $U(1)$-symmetry even if the target space is a HyperK\"{a}hler manifold. The axial $U(1)$ R-symmetry can only be enhanced to $SU(2)$ if the radius of this circle is zero, which happens to be the 2d theory on the same target.

From the analysis above of the gauge theory, it is not surprising that the K-theoretic vacuum equations are the Bethe ansatz equations of the XXZ model with the Hamiltonian
\begin{equation}\label{XXZ}
  H=\sum^{N}_{i=1}\left(S_{+,i+1}S_{-,i}+S_{-,i+1}S_{+,i}+2\Delta S_{z,i+1}S_{z,i}  \right),
\end{equation}
where the anisotropy $\Delta=\frac{1}{2}\left(Q+Q^{-1}\right)=\cos\left(\pi\widetilde{R}\right)$. So $\pi\widetilde{R}$ serves as the quantum deformation parameter $\hbar$ also known as $Q=e^{i\hbar}$. The 2d limit $R\mapsto 0$ gives $\Delta=1$, which is indeed the XXX model. It would be interesting to study the quantum K-theory of $T^{\ast}$Grassmannian \cite{Koroteev:2017nab} from the point of view of the spin chain.

 \subsection{XY And XYZ Spin Chains}
As discussed in section \ref{Sec:2.3}, if we put the 4d ${\cal N}$=1 gauge theory on space-time $\mathbb{R}^{2}\times \mathbb{T}^{2}$ with a generic torus $\mathbb{T}^{2}$, the axial R-symmetry is only a $\mathbb{Z}_{2}$ group. This reflects the fact the magnons' number is not conserved, rather, it is mod 2.

We argue that Eq.(\ref{4VEGR2}) is the Bethe equation for the XY model. To see this we first recall that the XX model has the Hamiltonian
\begin{equation}\label{XY1}
  H=\sum^{N}_{i=1}\left(S_{+,i+1}S_{-,i}+S_{-,i+1}S_{+,i}\right).
\end{equation}
Since the 4d gauge theory still has the ${\cal C}$, ${\cal P}$, and ${\cal T}$ symmetries but with only a mod 2 conserved magnons. This suggests that we should modify the Hamiltonian to be
\begin{equation}\label{XYH2}
  H=\sum^{N}_{i=1}\left(S_{+,i+1}S_{-,i}+S_{-,i+1}S_{+,i}+\gamma\left(S_{+,i+1}S_{+,i}+S_{-,i+1}S_{-,i}\right) \right),
\end{equation}
where $\gamma\neq 0$. This is indeed the XY model we are familiar with. Similarly, Eq.(\ref{4VEGR}) is the Bethe equation for the XYZ model with a Hamiltonian
\begin{equation}\label{XYZH}
  H=\sum^{N}_{i=1}\left(S_{+,i+1}S_{-,i}+S_{-,i+1}S_{+,i}+\gamma\left(S_{+,i+1}S_{+,i}+S_{-,i+1}S_{-,i}\right)+2\Delta S_{z,i+1}S_{z,i} \right).
\end{equation}

We end this section by commenting that no XYX or XYY model is associated with the 4d gauge theory on $\mathbb{R}^{2}\times \mathbb{T}^{2}$. Because if we had such an XYX model, the $U(1)$ symmetry that rotates the $xz$-plane has no physical origin. Furthermore, it contradicts the $\mathbb{Z}_{2}$ symmetry, which says that the parameters should satisfy $\Delta\neq 1\pm\gamma$, or $J_{x}\neq J_{y}\neq J_{z}$ where
\begin{equation}\label{}
  J_{x}=2(1+\gamma),\quad J_{y}=2(1-\gamma), \quad J_{z}=2\Delta.
\end{equation}

\subsection{A Remark On Some Other Models}
In \cite{Nekrasov:2009rc}, Nekrasov and Shatashvili found that the periodic Toda chain and elliptic Calogero-Moser system can be constructed from 4d ${\cal N}=2$ in $\Omega$-background reviewed in section \ref{Sec:2.4}. Their formulations rely solely on the bosonic variables of gauge theory. Thus, we expect that one can reformulate them in terms of Heisenberg spin chains, although we will not provide any detail in this paper. However, we want to point out that the complexified Hamitionians of these two systems have a natural explanation in our setup: Because the scatter factors of these two models are not pure phases, which indicates that the time-reversal is not a symmetry in these models.

So far, we have constructed several spin chains from gauge theories. In the next section, we will offer a deeper understanding of gauge theories from the perspective of the spin chain.

 \section{Duality As Symmetry And A Unification Of Gauge Theories}\label{Sec:4}

 In previous sections, we investigated how the Heisenberg spin chain emerges from a supersymmetric gauge theory in low energies. In this section, we take a reverse direction to address the question: what can we say about the gauge theory from the Heisenberg spin chain?

 Let us first recall what we have already discussed in two-dimensional gauge theory. In the far infrared, we should integrate out matters in gauge theory, which reduces it to an effective theory on the Coulomb branch with a twisted effective superpotential $\widetilde{W}_{{\rm eff}}$. The roles of charged matters have been changed in this process: They are asymptotic free particles in UV while they become solitons in the infrared. The fundamental piece is the domain wall $\Phi$ and the anti-domain wall ${\bar \Phi}$. All other BPS spectra are composite operators of these fundamental domain walls\footnote{ This is only true for flavor symmetries to be simply-connected compact Lie groups. However, a modified statement can be used for a general situation by taking into account the so-called higher symmetries.}. In order to make them gauge-invariant, one should attach them with Wilson lines. It is certainly possible to study the perturbative spectrum around one specific vacuum. In a region where the kinetic term of vector multiplet is suppressed\footnote{At the NLSM scale $\Lambda\ll\mu\ll e\sqrt{r}$ , where $e^{2}$ goes to infinity, the kinetic term can be certainly ignored. However, a dynamical gauge field can emerge in the IR with the effective gauge coupling $\sim\Lambda$. To suppress the kinetic term in the IR, one should consider the physical scale at $\mu\ll \Lambda$.}, one can find the gauginos are constraints of the system while the diagonal bosonic components can be expressed in terms of the matters fields by equations of motion of vector multiplet
 \begin{equation}\label{}
   A_{\mu}=\frac{i}{2}\frac{\sum^{N}_{j=1}\left({\bar \phi}_{j}\partial_{\mu}\phi_{j}-\partial_{\mu}{\bar \phi}_{j}\phi_{j}\right)}{\sum^{N}_{i=1}\mid \phi_{i}\mid^{2}},\quad\quad \sigma=-\frac{\sum^{N}_{j=1}{\bar \psi}_{+;j}\psi_{-;j}}{\sum^{N}_{i=1}\mid \phi_{i}\mid^{2}}.
 \end{equation}
   The other part is the off-diagonal components of the gauge field, which we call the $W$-bosons $W_{ab}$. They are ${\bar \phi}_{a}\phi_{b}$ classically. However, they can be expressed as $\widetilde{\phi}_{a}\phi_{b}$ in the exact quantum theory, where $\widetilde{\phi}$ are minus charged matters with the vector R-charge 2. This $\widetilde{\phi}$ field can be regarded as an anti-domain wall. All these tell us that the perturbative spectra are \textit{composite} operators of (anti-)domain walls.

The Heisenberg spin chain emerges in two-dimensional gauge theory at the intermediate scale, where the dynamics of perturbative degrees of freedom have been truncated out. The fundamental domain wall $\Phi_{i}$ and the anti-domain wall ${\bar \Phi}_{i}$ can be expressed in terms of the operators
\begin{equation}\nonumber
  {\cal D}_{i}\ {\rm and}\ {\bar{\cal D}}_{i},
\end{equation}
respectively, in the spin chain. Since other BPS spectra can be built from the fundamental ones, thus they can also be expressed in terms of the composite of spin operators in the spin chain.  Because the rank of the gauge group, $k$, corresponds to $k$ sites with spin-up excitation on a spin chain. As a gauge theory, the $U(k)$ group looks different from the gauge theory with the $U(k')$ theory if $k\neq k'$. For example, their perturbative spectrum is not the same. However, there could be a connection between the two gauge theories in low energy. It is the so-called Seiberg(-like) duality \footnote{The first several two-dimensional Seiberg dualities found in the literature are called Hori dualities \cite{Hori:2011pd}. In this paper, we use the terminology Seiberg duality originally found in four-dimensional to represent all them.}: they are different in the UV but identical in infrared. The simplest example is
\begin{equation}\label{}
  {\rm  Gr}(k;N)={\rm  Gr}(N{-}k;N).
\end{equation}
The first gauge theory is $U(k)$ gauge theory with $N$ fundamental matters, while the second one is $U(N-k)$ gauge with $N$ fundamentals.
Of course, there are more complicated examples \cite{Hori:2011pd}. But the idea is the same. So what does the duality correspond to in a spin chain? We notice that the duality between two gauge theories can be formalized by a map in the spin chain:
\begin{equation}\label{DMISPC}
  k\  {\rm  spin{-}up}\ {\rm sites} \mapsto  N{-}k\  {\rm  spin{-}up\ sites}.
\end{equation}
 This map is actually the ${\cal P}$ symmetry when $\widetilde{q}=\pm 1$. To see this, we recall that ${\cal P}$ acts on sites as
 \begin{equation}\label{}
   {\cal P}{\bar \lambda}_{\pm} {\cal P}^{-1}={\bar \lambda}_{\mp},\quad{\rm and }\quad {\cal P}S_{z} {\cal P}^{-1}=-S_{z},
 \end{equation}
so it indeed defines the map (\ref{DMISPC}). However, even if $\widetilde{q}\neq\pm 1$, one can still apply ${\cal P}$ to build the map (\ref{DMISPC}) by associating with a change of the Hamiltonian as well:
 \begin{equation}\label{}
   h_{1}\mapsto {\bar h}_{1}.
 \end{equation}

 The duality is extremely nontrivial in quantum field theory since two different ones are connected by the dual transformation, which also means that the duality, in general, is not a symmetry in field-theoretic language. Thus, it is not \textit{manifest}. However, this phenomenon is a consequence of ``symmetry" in the spin chain, so it is \textit{manifest} in this context. Furthermore, because it is a symmetry in a closed spin chain, which can be explicitly broken by turning on the background magnetic field $B$ with a coupling
 \begin{equation}\label{}
   B\sum^{N}_{i=1} S_{z,i}
 \end{equation}
 in the Hamiltonian. However, one may still treat ${\cal P}$ as symmetry in this case if one imposes the operation ${\cal P}$ on the magnetic field as well by
 \begin{equation}\label{}
   B\mapsto -B.
 \end{equation}
 Promoting the background field to be a dynamical field, in fact, has been widely used in quantum field theory.

Following the above philosophy, we can further propose that different rank gauge theories can be unified into a single system: the Heisenberg spin chain. To see this, we first notice that the operator $\sum^{N}_{i=1}S_{+,i}$ acts on the spin chain by mapping
 \begin{equation}\label{}
   k\ {\rm spin{-}up\ sites}\mapsto k{+}1\ {\rm spin{-}up\ sites}.
 \end{equation}
  It interprets as a map from the $U(k)$ gauge theory to the $U(k+1)$ one on the gauge theory side. More precisely, at the intermediate scale, the $U(k)$ gauge theory can map to the $U(k+1)$ one. Certainly, understanding the counterpart of the operator $\sum^{N}_{i=1}S_{+,i}$ in gauge theory would be an interesting question. We expect that it can only be defined in the infinity of the field space since it not only changes the gauge group but also bring more vacua to the system. On the other hand, the operator $\sum^{N}_{i=1}S_{-,i}$ defines a map in the opposite direction
\begin{equation}\label{}
   k+1\ {\rm spin{-}up\ sites}\mapsto k\ {\rm spin{-}up\ sites}.
 \end{equation}
 With these ingredients, we claim that each rank of gauge group with the same kind of representation is actually a sub-space of a big theory. For example, we have a long sequence:
\begin{equation}\label{}
  Gr(0;N)\xrightarrow{\sum^{N}_{i=1}S_{+,i}}Gr(1;N)\xrightarrow{\sum^{N}_{i=1}S_{+,i}}\cdots\xrightarrow{\sum^{N}_{i=1}S_{+,i}}Gr(N;N).
\end{equation}
And the inverse one:
\begin{equation}\label{}
Gr(0;N)\xleftarrow{\sum^{N}_{i=1}S_{-,i}}Gr(1;N)\xleftarrow{\sum^{N}_{i=1}S_{-,i}}\cdots\xleftarrow{\sum^{N}_{i=1}S_{-,i}}Gr(N;N).
\end{equation}
The physical reason for these two sequences is that they share the same global symmetries, even though their gauge groups are different. Before ending this section, we point out again that our claim is only verified at the intermediate scale, where an emergent spin chain appears. However, our claim may also be correct at higher energies for the BPS spectra, which relies on the exact results of gauge theory discussed in section \ref{Sec:2}.

 \section{The Yang-Baxter Equation}\label{Sec:5}
In previous sections, we have shown how to construct a Heisenberg spin chain from the supersymmetric gauge theory. However, it has other formulations for describing the dynamics of an integrable system. For example, the famous Yang-Baxter equation is a starting point. This section is devoted to investigating the Yang-Baxter equation with new insight.

 Let us consider first the scattering situations where particles preserve their momentum while changing their internal states, so it is an integrable system.  Assume these particles' internal quantum numbers take values in some vector space ${\cal V}$. A particle is parameterized by (exponential of) momentum or rapidity denoted by the complex/spectral parameter $p$.

\begin{figure}[h!]{}
\centering
\includegraphics[width = 0.4 \columnwidth]{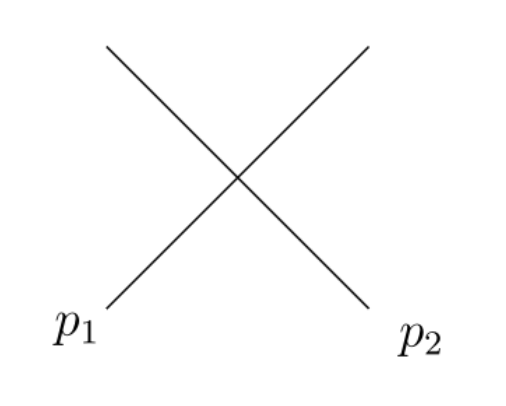}
\caption{Scattering of two particles: their spectral parameters $p_{1}$ and $p_{2}$ are unchanged but their ``internal" vector space is transformed.}
\label{fig:1}
\end{figure}

 For the scattering of two particles (Fig.\ref{fig:1}) denoted as $a$ and $b$, we then define the so-called $R$-matrix as $R_{ab}(p_{a}- p_{b})$ : ${\cal V}_{a}\otimes {\cal V}_{b}\mapsto{\cal V}_{a}\otimes {\cal V}_{b}$, where we have assumed the Lorentz symmetry. In the context of the present paper, the ``particle" can be understood as a composition of two sites labeled by $\sigma_{a}'s$, or
 \begin{equation}\label{}
   \frac{\sigma_{a}+\frac{i}{2}}{\sigma_{a}-\frac{i}{2}}.
 \end{equation}
 We may choose the site with a vanishing momentum as the reference site, or we can pick an auxiliary one to do the job. In this sense, we have
  \begin{equation}\label{}
   p_{a}=\frac{1}{i}\log\sigma_{a}\quad {\rm or} \quad p_{a}=\frac{1}{i}\log\left(\frac{\sigma_{a}+\frac{i}{2}}{\sigma_{a}-\frac{i}{2}}\right).
\end{equation}
 For most cases in our paper, this means that we have the basis of ${\cal V}$ to be
 \begin{equation}\label{}
   \mid+,+\rangle,\quad \mid-,+\rangle,\quad \mid+,-\rangle,\quad \mid-,-\rangle.
 \end{equation}
  Since the basis is transformed under the $R$-matrix, we expect that the off-diagonal entries of the $R$-matrix correspond to the (anti)-domain walls in our context.

   To see this, we first consider the spin chain for Grassmannian. Since this system does not have the time reversal symmetry for a generic $\widetilde{q}$, the Hamiltonian discussed in section \ref{GRXX} is chosen to be the complex one. Thus, there are two possibilities of
  \begin{equation}\label{}
  \left(
    \begin{array}{cccc}
      a &  &  &  \\
       & b & c &  \\
       & \widetilde{c} &\widetilde{ b }&  \\
       &  &  &  \widetilde{a}\\
    \end{array}\right)
  \end{equation}
  with the matrix entries, given in \cite{Gorbounov:2014}:
  \begin{equation}\label{}
 \begin{tabular}{|c|c|c|c|c|c|c|}
   \hline
    & a & b & c & $\widetilde{c}$ & $\widetilde{b}$ & $\widetilde{a}$ \\
   $R(v_{a},v_{b})$ & 1 &  1 & 0 &$ v_{b}\ominus v_{a}$ & 1+$\beta v_{b}\ominus v_{a}$ & 1 \\
   $\widetilde{R}(v_{a},v_{b})$ & 1 & 1 & $v_{a}\ominus v_{b}$&  0 & 1+$\beta v_{a}\ominus v_{b}$ & 1 \\
   \hline
 \end{tabular},
  \end{equation}
where
\begin{equation}\label{}
  v_{a}\ominus v_{b}=\frac{v_{a}-v_{b}}{1+\beta v_{b}}.
\end{equation}
When $\beta=0$ and take $v_{a}=\sigma_{a}$, it describes the integrable system for quantum cohomology of Grassmannian. When $\beta=-1$ and setting $v_{a}=z_{a}$, the system captures the quantum K-theory of Grassmannian. Since $c$ or $\widetilde{c}$ is vanishing, this model is the so-called five-vertex model. Our paper provides a physical explanation for this. The complex Hamiltonian in the spin chain only includes a domain wall or anti-domain wall for a generic $q=e^{-t}$ since it lacks the time-reversal symmetry except at $\widetilde{q}=\pm 1$ for quantum cohomology. The entry $c$ describes the fluctuation
\begin{equation}\label{}
  \mid+,-\rangle\mapsto \mid-,+\rangle,
\end{equation}
which is exactly the anti-domain wall configuration. So if we are considering the Hamiltonian
\begin{equation}\label{}
  H=\sum_{i} {\cal D}_{i}+f({\cal D}_{i})+\textbf{I},
\end{equation}
the entry $c=0$. Similarly, if we consider the conjugate Hamiltonian ${\bar H}$, we have $\widetilde{c}=0$.

 In XXX and XXZ models, the Hamiltonian is a hermitian operator, which says the entries satisfy
 \begin{equation}\label{}
   a=\widetilde{a},\quad\quad  b=\widetilde{b},\quad\quad c=\widetilde{c}.
 \end{equation}
 They correspond to the six-vertex models in the integrable statistical mechanics. While in the XYZ model, besides the usual domain wall configuration, we also have the (anti)-domain wall configuration:
  \begin{equation}\label{}
     \mid-,-\rangle\mapsto \mid+,+\rangle,\quad {\rm or} \quad \mid+,+\rangle\mapsto \mid-,-\rangle.
  \end{equation}
  Thus, it is an eight-vertex model with the $R$-matrix
    \begin{equation}\label{}
  \left(
    \begin{array}{cccc}
      a &  &  & d \\
       & b & c &  \\
       & c & b &  \\
      d &  &  &  a\\
    \end{array}\right).
  \end{equation}
Symmetries play a crucial role in finding the connections between gauge theory and $R$-matrices. However, one can define a general $R$-matrix abstractly, and it would be interesting to understand its counterpart in quantum field theory. For example, the Lorentz symmetry of the two-dimensional gauge theory would be broken by turning on a $\Omega$-deformation to the two-dimensional space-time, which may correspond to a more general $R$-matrix.

\begin{figure}[h!]{}
\centering
\includegraphics[width = 0.4 \columnwidth]{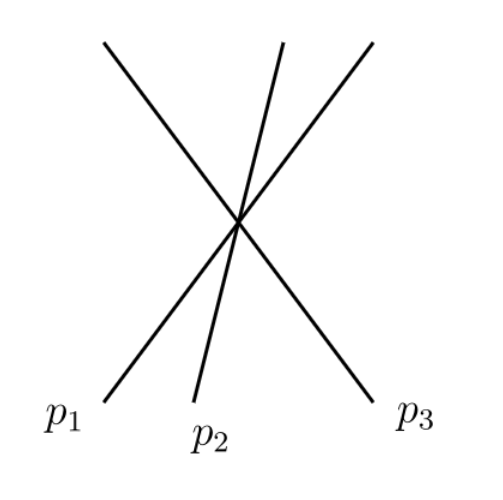}
\caption{Three particles cross at the same point.}
\label{fig:2}
\end{figure}

\begin{figure}[h!]{}
\centering
\includegraphics[width = 0.6 \columnwidth]{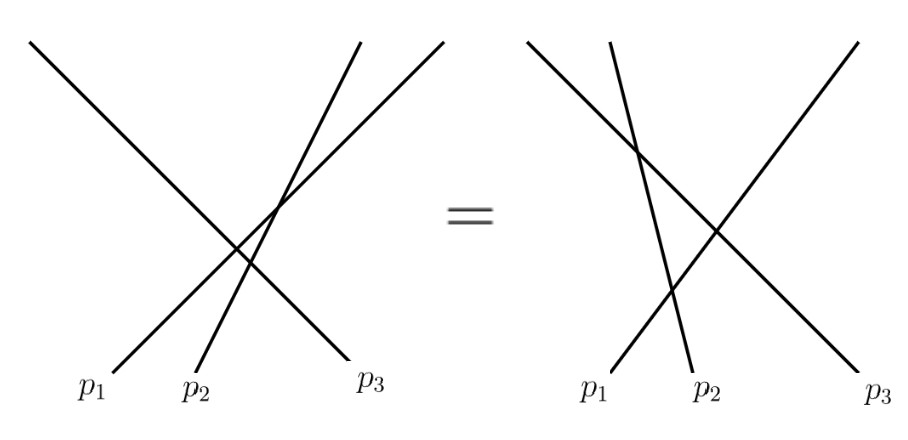}
\caption{The Yang-Baxter equation arises from the equivalence between these two pictures.}
\label{fig:3}
\end{figure}

Now, we consider a scattering process with three ``particles". It can be denoted mathematically as the map
\begin{equation}\label{}
  {\cal V}_{a}\otimes {\cal V}_{b}\otimes {\cal V}_{c}\mapsto{\cal V}_{a}\otimes {\cal V}_{b}\otimes {\cal V}_{c}.
\end{equation}
We notice that the map
\begin{equation}\label{}
   {\cal V}_{a}\otimes {\cal V}_{b}\mapsto{\cal V}_{a}\otimes {\cal V}_{b}
\end{equation}
  has components that correspond to domain wall configurations. Since we consider the cases where the vacua are isolated. Thus, there is no room for a genuinely three-indices operator $R_{abc}$ (Fig.\ref{fig:2}), and no higher-indices operators either.
   For example, the map
  \begin{equation}\label{}
    \mid+,-,+\rangle\mapsto \mid-,+,-\rangle
  \end{equation}
is not allowed in our consideration. So all scattering processes are decomposed into a multiplication of the scattering between two particles. The scattering amplitude of three particles shall be well-defined, which means the amplitude does not depend on the ordering in the multiplication. From this constraint, one will derive the well-known Yang-Baxter equation (Fig.\ref{fig:3})
\begin{equation}\label{}
  R_{12}(u_{1},u_{2})R_{13}(u_{1},u_{3})R_{23}(u_{2},u_{3})=R_{23}(u_{2},u_{3}) R_{13}(u_{1},u_{3})R_{12}(u_{1},u_{2}).
\end{equation}
 So the existence of the Yang-Baxter equation has a dynamic reason in our context. However, we do not exclude the possibility in other supersymmetric gauge theories that causes the interaction in Fig.\ref{fig:2}. If it is further an integrable system, the Yang-Baxter equation could be generalized for these interactions. Of course, one could assume the abstract Yang-Baxter equation first and then find all possible physical solutions. This program is also fruitful\cite{Kulish:1980ii}.

 Finally, we end this section by mentioning that the integrability of the dynamics of domain walls has also been investigated in \cite{Fendley:1990zj,Fendley:1992dm} for abelian theories in other contexts.

 \section{Comment On Four-dimensional Chern-Simons Theory}\label{Sec:6}
It was studied in \cite{Costello:2017dso, Costello:2013sla, Costello:2013zra} that the four-dimensional Chern-Simons gauge theory can describe some integrable systems. The setup in question is only defined on a special four-manifold with the structure as a product of Riemann surfaces, $M=\Sigma\times C$, where $\Sigma$ is a smooth oriented 2-manifold and $C$ is a complex manifold endowed with a holomorphic (or meromorphic) 1-form $\omega$. The space-time coordinate is denoted by $(x,y,z,{\bar z})$. The theory only has the gauge field that is an adjoint representation of the complex Lie group $G$ with the component fields as $A=A_{x}dx+A_{y}dy+A_{z}dz+A_{\bar z}d{\bar z}.$ The action is
 \begin{equation}\label{}
   I=\frac{1}{2\pi}\int_{M}\omega\wedge CS(A),
 \end{equation}
 where CS(A) is the Chern-Simons three-form
 \begin{equation}\label{}
   {\rm CS(A)}:={\rm Tr}\left(A\wedge dA+\frac{2}{3}A\wedge A\wedge A\right).
 \end{equation}
 From the action, one can find the theory has an extra gauge symmetry: $A\rightarrow A+ \varphi dz$. This gauge symmetry can be fixed by choosing $A_{z}=0$. For the theory to be defined perturbatively without introducing essentially new ingredients, we require the 1-form $\omega$ has no zero. With this constraint, there are only three possibilities for $C$: (1) $C=\mathbb{C}$, $\omega=dz$, double poles at $\{\infty\}$, XXX model; (2) $C=\mathbb{C}^{\star}$, $\omega=\frac{dz}{z}$, poles at $\{0, \infty\}$, XXZ model; (3) $C=E=\mathbb{C}/\left(\mathbb{Z}+\tau\mathbb{Z}\right)$, $\omega=dz$, no poles, XYZ model. The authors in \cite{Costello:2017dso} have computed the three possible $R$-matrix for various gauge groups and further observed that
 \begin{equation}\label{}
   R^{{\rm XYZ}}_{\hbar,\tau}\left(v\right)\xrightarrow{\tau\rightarrow i\infty} R^{{\rm XXZ}}_{\hbar}\left(v\right)\xrightarrow{\hbar\rightarrow 0} R^{{\rm XXX}}_{\hbar}\left(v\right).
\end{equation}

 On the other hand, our construction relies on the twisted effective superpotential and domain walls. The connection between these two different constructions is an important question. We first list some similarities between them. Let us recall the 4d ${\cal N}$=1 gauge theory on space-time $\mathbb{R}^{2}\times C$, we have: (1) If $C=\mathbb{T}^{2}$, it gives an XYZ spin chain; (2) Taking $\tau\mapsto i\infty$, it reduces to an XXZ spin chain; (3) Taking $\widetilde{R}=\frac{\hbar}{\pi}\rightarrow 0$ further, it becomes an XXX spin chain. Therefore, the same $C$ is used in the two routes to integrable systems.

 Although the above similarities are encouraged, there are many crucial differences: The theory defined on the Coulomb branch has a manifest ${\cal N}$=(2,2) supersymmetry and only maintains an abelian-like gauge symmetry: $T\rtimes S_{G}$, where $T$ and $S_{G}$ are the maximal torus and the Weyl symmetry of $G$, respectively. By contrast, there is no manifest supersymmetry in the 4d CS theory, and it preserves the gauge symmetry $G$. This issue could be solved as follows: We first integrate the massive fermions out in the twisted effective superpotential, and it reduces to the $T/T$ gauged WZW model. Based on the observation of the correspondence between a $G/G$ gauged WZW model and its associated $T/T$ one in \cite{Blau:1993tv}, we claim that
 \begin{equation}\label{STM1}
   G/G \ {\rm gauged\ WZW \ model}\equiv \left(\left(C^{\star}\right)^{{\rm Rank} G}/S_{G}, \widetilde{W}_{{\rm eff}}\right)
 \end{equation}
 is true at least for some cases. It was further observed in \cite{Blau:1993tv} that
 \begin{equation}\label{STM2}
   {\rm CS\ theory}\ {\rm on}\ \Sigma\times S^{1}\equiv  G/G \ {\rm gauged\ WZW \ model}\ {\rm on}\ \Sigma.
 \end{equation}
 However, the above statements have only been proved for cases such as $U(k)$ gauge group with fundamental matters \cite{Witten:1993xi} and $U(k)$ gauge group with $N$-fundamental fields plus an adjoint matter \cite{Gukov:2015sna}. But, in our situation, we also need to include extra $N$ anti-fundamental matters. The verification of statements (\ref{STM1}) and (\ref{STM2}) for our case is still missing.

 The second difference is that the gauge group $G$ in 4d CS theory is a complex semi-simple Lie group, while the gauge group in ${\cal N}$=(2,2) theory is real. Although some properties of 2d $G/G$ gauged WZW can be extended to the $G_{\mathbb{C}}/G_{\mathbb{C}}$ one \cite{Witten:1993xi,Blau:1993tv}, a complete understanding is still missing. Finally, we want to mention that Gukov and Witten in \cite{Gukov:2008ve} stated a new perspective on the quantization of Chern-Simons gauge theory. A crucial procedure in their setup is to embed the Chern-Simons theory into the one with a complex gauge group $G_{\mathbb{C}}$. We expect this new insight helps us understand the connection between gauge theories and integrable systems, which we leave a detailed investigation to future work.

\begin{center}
\section*{Acknowledgement}
\end{center}
We would like to thank V. Gorbounov, S. Gukov, A. Klemm, C. Korff, N. A. Nekrasov, S. L. Shatashvili, and E. Witten for email correspondence. Research of Wei Gu was supported in part by NSF grant PHY-1720321.
\\

\end{document}